\documentclass[11pt,letterpaper]{article}
\usepackage[T1]{fontenc}
\IfFileExists{lmodern.sty}{\usepackage{lmodern}\usepackage{microtype}}{\usepackage[expansion=false]{microtype}}
\usepackage[margin=1in]{geometry}
\usepackage{amsmath,amssymb,amsthm,mathtools}
\usepackage{booktabs,tabularx,enumitem,needspace,array}
\usepackage[hidelinks]{hyperref}
\usepackage[capitalize,nameinlink,noabbrev]{cleveref}
\setlist[itemize]{leftmargin=*,itemsep=2pt,topsep=4pt}
\setlist[enumerate]{leftmargin=*,itemsep=2pt,topsep=4pt}
\allowdisplaybreaks[1]
\newtheorem{theorem}{Theorem}[section]
\newtheorem{lemma}[theorem]{Lemma}
\newtheorem{corollary}[theorem]{Corollary}
\newtheorem{proposition}[theorem]{Proposition}
\theoremstyle{definition}

\newcommand{\F}{\mathbb F_2}
\newcommand{\D}{D}
\newcommand{\depth}{\operatorname{cert}}
\newcommand{\Max}{\operatorname{MaxSupp}}

\newcommand{\aff}{\operatorname{aff}}
\newcommand{\spanf}{\operatorname{span}}
\newcommand{\Ind}{\operatorname{Ind}}
\newcommand{\poly}{\operatorname{poly}}
\newcommand{\Sig}{\Sigma_2^{\oplus\mathrm P}}

\newcommand{\KarthikAffiliation}{Independent Researcher}
\newcommand{\NicholasAffiliation}{Department of Computer Science, Stony Brook University}
\newcommand{\KarthikEmail}{\textcolor{black}{karthiksheshadri@gmail.com}}
\newcommand{\NicholasEmail}{\textcolor{black}{nicholas.smirnov@stonybrook.edu}}
\usepackage{xcolor}

\hypersetup{pdftitle={The Complexity of Choosing Parity Projections},
 pdfauthor={Karthik Sheshadri and Nicholas Smirnov}}

\begin{document}
\hypersetup{pageanchor=false}
\begin{titlepage}
\centering
\vspace*{0.1in}
{\LARGE\bfseries The Complexity of Choosing\par Parity Projections\par}
\vspace{0.2in}
{\large Karthik Sheshadri$^{1}$ and Nicholas Smirnov$^{2}$\par}
\vspace{6pt}
{\normalsize $^{1}$\KarthikAffiliation, \texttt{\KarthikEmail}\par}
\vspace{2pt}
{\normalsize $^{2}$\NicholasAffiliation, \texttt{\NicholasEmail}\par}
\vspace{0.12in}
{October 2026\par}
\vspace{0.12in}
\begin{abstract}
\normalsize
Cube attacks in cryptanalysis seek small sets of Boolean variables
(cubes) whose parity sum is constant. We study the bounded constant
derivative problem introduced by Smirnov (2026), which seeks a smallest
cube with sum 1, and its sum-0 analogue. Despite their similarity, the
problems differ sharply in parameterized complexity: the sum-1 problem
admits a $2^{O(k^2)}\operatorname{poly}(N)$-time algorithm, where $k$
is cube size, whereas the sum-0 problem is not fixed-parameter
tractable unless $\mathrm{FPT}=\mathrm{W}[1]$. Without a bound on cube
size, both problems are $\Sigma_2^{\oplus\mathrm{P}}$-complete and
NP-hard to approximate within $N^{1-\varepsilon}$ for every fixed
$0<\varepsilon<1$.

Our hardness proofs use a crown construction that embeds maximal
monomials of parity-projected circuits into positive 2CNF. This yields
completeness and inapproximability results, as well as a dichotomy for
Boolean constraint languages of arity at most two: the bounded constant
derivative problem is in P when all relations are affine and
$\Sigma_2^{\oplus\mathrm{P}}$-complete otherwise. We also give
algorithms parameterized by the number of quadratic rounds, treewidth,
or ordered branching-program width, with lower bounds for all three
unless $\mathrm{NP}\subseteq\mathrm{SUBEXP}$. The treewidth lower bound
holds for circuits; the case of 2CNF remains open.
\end{abstract}
\end{titlepage}
\setcounter{page}{1}
\hypersetup{pageanchor=true}

\section{Introduction}

Let $f$ be a Boolean function of the variables $x_1,\dots,x_n$ and let
$S$ be a nonempty set of these variables. Fixing the variables outside
$S$ to a value $y$ and summing $f$ over all $2^{|S|}$ assignments to the
variables in $S$ gives, modulo two, a Boolean function of $y$,
\[
 (\D_Sf)(y)=\bigoplus_{z\in\{0,1\}^S}f(z,y).
\]
Cryptanalysts call $S$ a cube and $\D_Sf$ its
superpoly~\cite{dinurshamir2009}.

The bounded constant derivative problem (BCD) of~\cite{smirnov2026}
asks, given $f$ as a formula or a circuit and a number $k$, whether some
set $S$ of at most $k$ variables has $\D_Sf\equiv1$. We study BCD and its
constant-zero counterpart, which asks the same question for
$\D_Sf\equiv0$.

Following Smirnov~\cite{smirnov2026}, we call a set $S$ with
$\D_Sf\equiv1$ a derivative certificate of $f$ (certificate for short)
and write $\depth(f)$ for the size of the smallest one. We call a set $S$ with $\D_Sf\equiv0$ a zero cube and write $\zeta(f)$ for the size of the smallest one. Both quantities are $+\infty$ when no set exists.

The problem of the constant superpoly is well-motivated in
cryptanalysis. Dinur and Shamir~\cite{dinurshamir2009} showed that a
cube whose superpoly is a simple function of the secret key yields an
equation in the key bits. Sufficiently many such cubes recover the key
of a reduced round stream cipher. For any small cube, a random
polynomial is extremely unlikely to have a constant superpoly; for a
uniformly random function of $n$ variables and a cube of size $k$, the
probability is $2^{1-2^{n-k}}$. The problem involves searching for a
cube with a constant superpoly given the function, which is provided
either explicitly as a circuit or as a composition of low degree round
maps.

In~\cite{drs2024}, Dutta, Rajasree and Sarkar address the question of
how hard this search is in the worst case. They show that monomial
prediction is $\oplus\mathrm P$-complete. Since every coefficient above
the chosen set must be controlled and the set itself must be found, our
problem is more demanding than monomial prediction.

Smirnov~\cite{smirnov2026} observed that both conditions have a simple
description in terms of the algebraic normal form of $f$. $S$ is a
certificate exactly when $x_S$ is a term of $f$ that is contained in no
larger term, and it is a zero cube exactly when no term of $f$ contains
$x_S$. Therefore, a certificate is a small maximal term and a zero cube
is a small set that no term covers. When the terms of $f$ are listed,
both are easy to find. However, a short formula may have exponentially
many terms, and the description does not by itself give an algorithm.
Smirnov showed that BCD is coNP-complete on circuits for each fixed $k$,
and that for variable $k$ it belongs to $\Sig$. Smirnov also posed an
open question, namely, whether BCD is complete for $\Sig$. A preliminary
manuscript by Sheshadri~\cite{sheshadri2026} proved hardness results for
circuits and for restricted $3$CNF formulas. To the best of our
knowledge, nothing in published literature addressed formulas with two
literals per clause, the zero cube version, or the parameterized
complexity of either problem.

Our proofs evolve from three lineages of prior work.
Faben~\cite{faben2008} introduced reductions that preserve the number of
solutions of a generalized satisfiability problem modulo $k$, and used
them to classify the complexity of counting those solutions modulo $k$.
His work was developed further by Kazeminia and
Bulatov~\cite{kazeminia2025}, who extended modular counting to general
constraint problems and reduced them to problems with binary
constraints only. However, note that a minimization problem is
sensitive to the maximal terms of the resulting formula, which related
work does not address.

Approximation bounds of the same kind appear in
Umans~\cite{umans1999,umans2001}, who proved strong inapproximability
results for $\Sigma_2$ minimization problems such as the shortest
implicant.

Whereas for algorithms based on structural width the closest results
are those for quantified satisfiability~\cite{lampismitsou2017} and
projected model counting~\cite{fichte2023}, where the summed variables
are given, our problems require the set of summed variables to be part
of the solution.

\begin{table}[t]
\centering
\footnotesize
\renewcommand{\arraystretch}{1.15}
\begin{tabularx}{\textwidth}{@{}>{\raggedright\arraybackslash}p{0.31\textwidth}>{\raggedright\arraybackslash}X>{\raggedright\arraybackslash}X@{}}
\toprule
Input, and what is asked & Certificates ($\D_Sf\equiv1$) & Zero cubes ($\D_Sf\equiv0$)\\
\midrule
linear system $Hx=b$, exact & P (column basis) & NP-hard (minimum distance)\\
signed $2$CNF, cube size $\le k$ & $2^{O(k^2)}\poly(\ell)$, kernel of $2^{O(k)}$ bits & coW[1]-hard to tell $\le k$ from $>g(k)$, any computable $g$\\
positive $2$CNF, exact & $\Sig$-complete & $\Sig$-complete\\
positive $2$CNF, poly-time approximation & $N^{1-\varepsilon}$ and $\ell^{1-\varepsilon}$ hard & $N^{1-\varepsilon}$ and $\ell^{1/2-\varepsilon}$ hard\\
$3$CNF, each fixed $k$ & P, but Horn $3$CNF is coW[1]-hard in $k$ & coNP-complete\\
$4$CNF, each fixed $k$ & coNP-complete & coNP-complete\\
\midrule
$r$ quadratic rounds, $n$ inputs & $n^{O(2^r)}\poly(\ell)$, no $n^{2^{o(r)}}\poly(\ell)$ & not treated here\\
treewidth $w$ (circuits) & $2^{2^{O(w)}}\poly(\ell)$, tight & same algorithm\\
ordered branching programs, width $w$ & $2^{O(w^2)}n$, no $2^{w^{o(1)}}\poly(\ell)$ & same algorithm\\
\bottomrule
\end{tabularx}
\caption{Summary of results. Approximation entries hold for every fixed
$\varepsilon>0$ unless P=NP; for certificates on positive $2$CNF they hold
under the weaker assumption $\Sig\ne\mathrm P$. The lower bounds in the last three rows hold unless
$\mathrm{NP}\subseteq\mathrm{SUBEXP}$. The zero-cube hardness entries of the $2$CNF rows hold on
positive unit-free
$2$CNF that is satisfiable, has every variable essential, and comes with
a zero cube supplied. The first row is the example of \cref{sec:overview}.}
\label{tab:results}
\end{table}

\paragraph{Notation.}
Throughout, $N$ is the number of variables of the input and $\ell$ is
its encoding length. A signed $2$CNF may contain both positive
and negative literals, a positive $2$CNF contains only positive
literals, and a $2$CNF is unit-free if every clause has two
variables. FPT and W[1] are the standard notions of parameterized
complexity~\cite{cygan2015}. A problem is coW[1]-hard if it is at least
as hard as the complement of $k$-Clique, and such a problem is not
fixed-parameter tractable unless FPT=W[1]. SUBEXP is the class of
problems that can be solved in time $2^{\ell^{\varepsilon}}$ for every
fixed $\varepsilon>0$. The Exponential Time Hypothesis (ETH) asserts that
for some $\delta>0$, $3$SAT on $N$ variables cannot be solved in time
$2^{\delta N}$~\cite{cygan2015}. ETH was recently
refuted~\cite{ethfalse}. Our conditional lower bounds therefore assume
only that $\mathrm{NP}\not\subseteq\mathrm{SUBEXP}$, a weaker hypothesis
that remains open.

\paragraph{Our contributions.}
We determine the complexity of both problems on binary formulas.

\begin{theorem}\label{thm:main-separation}
\begin{enumerate}[label=(\alph*)]
\item On signed $2$CNF, a minimum certificate of size at most $k$ can be
found, or shown not to exist, in $2^{O(k^2)}\poly(\ell)$ time, and the
problem has a kernel with $2^{O(k)}$ bits.
\item On positive unit-free $2$CNF, for every computable function
$g(k)\ge k$, deciding whether $\zeta(f)\le k$ or $\zeta(f)>g(k)$ is
coW[1]-hard. The hard instances are satisfiable, every variable is
essential, and a zero cube is supplied with the instance.
\end{enumerate}
\end{theorem}

In part (a) we establish an FPT algorithm and a kernel for minimum
certificates on signed $2$CNF (\cref{thm:small-one} and \cref{thm:kernel}). Part
(b) concerns the exclusion of FPT algorithms that tell a small zero cube
from a large one (\cref{thm:zero-gap}). Note that the exclusion covers
any FPT algorithm that approximates the smallest zero cube within a
computable factor, even one whose running time depends on the true
optimum rather than on $k$ (\cref{cor:zero-approx}). Notice that the
classical classification does not distinguish the two problems. Both
are $\Sig$-complete on positive unit-free $2$CNF
(\cref{thm:hardness} and \cref{thm:zero-classical}), and are hard for the same
reason, as both reductions go through the lifting theorem described
below. The difference appears only when the cube is required to be
small.

For the purposes of polynomial time approximation, we show that neither
minimum can be approximated within $N^{1-\varepsilon}$ on positive
$2$CNF, for any fixed $0<\varepsilon<1$. In terms of $\ell$ the bounds
are $\ell^{1-\varepsilon}$ for certificates on positive $2$CNF and for
zero cubes on satisfiable $3$CNF, and $\ell^{1/2-\varepsilon}$ for zero
cubes on positive $2$CNF.

All of the classical hardness results come from a single construction,
the weighted lifting theorem of \cref{sec:lift} (\cref{thm:lift}). It
takes an arbitrary circuit, sums out some of its inputs, and produces a
positive $2$CNF whose certificates are exactly the maximal terms of the
summed circuit.

The same construction gives a dichotomy for Boolean constraint
languages of arity at most two. BCD is in P when all relations are
affine and $\Sig$-complete otherwise (\cref{thm:binary}).

Additionally, we contribute algorithms parameterized by the number of
quadratic rounds $r$ and by the treewidth $w$, running in time
$n^{O(2^r)}\poly(\ell)$ and $2^{2^{O(w)}}\poly(\ell)$ respectively, with
lower bounds, unless $\mathrm{NP}\subseteq\mathrm{SUBEXP}$, that rule out
$n^{2^{o(r)}}\poly(\ell)$ time in a stated range of $r$
(\cref{thm:rounds}) and $2^{2^{o(w)}}\poly(\ell)$ time for circuits
(\cref{thm:width}). For branching programs of width $w$ we give
a $2^{O(w^2)}n$ algorithm (\cref{thm:branching}), and under the same
assumption we rule out $2^{w^{o(1)}}\poly(\ell)$ time
(\cref{prop:obdd-hard}).

We note two boundaries. Clause width three already defeats the
algorithm of part (a), because on Horn $3$CNF certificates have the same
arbitrary parameterized gap that zero cubes have on positive $2$CNF
(\cref{thm:horn}). However, for each fixed $k$ the problem remains in P
on $3$CNF and becomes coNP-complete on $4$CNF (\cref{prop:fixed}).

The positive-$2$CNF hardness survives on formulas whose independence
complex strongly collapses to a point (\cref{app:topology}). These
results are listed in \cref{tab:results}.

The remainder of this paper is organized as follows. \Cref{sec:overview}
explains the ideas behind the three main proofs.
\Cref{sec:small-one,sec:zero,sec:lift} contain the proofs of part (a), of
part (b), and of the classical results, and \cref{sec:other} treats the
other representations. All deferred proofs are in the appendix.
\section{Proof summaries}\label{sec:overview}
In this section, we describe a central fact about derivatives that all
of our proofs rely on: summing $f$ over a set $S$ keeps exactly the
terms of $f$ that contain all of $S$, with the variables of $S$ removed,
and cancels every other term.

\subsection{The maximal-term rule}\label{sec:basic}
Every Boolean function has a unique representation as a multilinear
polynomial over $\F$,
\[
 f=\sum_Ac_Ax_A,\qquad x_A=\prod_{i\in A}x_i,\quad c_A\in\{0,1\},
\]
where $A$ ranges over the subsets of the variables. This is the
algebraic normal form of $f$, and the sets $A$ with $c_A=1$ are its
supports. Summing over a single variable $x_i$ either cancels a term or
removes $x_i$ from it: a term with $i\notin A$ takes the same value for
both settings of $x_i$ and cancels, while a term with $i\in A$ vanishes
at $x_i=0$ and loses the factor $x_i$ at $x_i=1$. Iterating over the
variables of $S$ gives the following.

\begin{lemma}[Smirnov~\cite{smirnov2026}]\label{lem:anf}
$\D_Sf=\sum_{A\supseteq S}c_Ax_{A\setminus S}$. Consequently, for nonempty
$S$, $\D_Sf\equiv1$ if and only if $S$ is a support of $f$ and no strict
superset of $S$ is one (we say that $S$ is a maximal support), and
$\D_Sf\equiv0$ if and only if no support of $f$ contains $S$.
\end{lemma}
\begin{proof}
Distinct supports containing $S$ remain distinct after $S$ is deleted,
so there is no cancellation on the right-hand side. The sum is the
constant $1$ exactly when the empty product is the only term that
survives, and it is $0$ exactly when no term survives.
\end{proof}

We denote the family of maximal supports of $f$ with $\Max(f)$. We
maintain the conventions $\Max(0)=\varnothing$ and
$\Max(1)=\{\varnothing\}$. For example, consider: $f=x_1x_2+x_3$. The set
$\{3\}$ is a certificate because $x_3$ is a term that no other term
extends, and $\{1,3\}$ is a zero cube because no term contains both $x_1$
and $x_3$.

Our remaining arguments reuse the following key consequences of this
lemma. Firstly, zero cubes are closed under the addition of variables;
whereas certificates are not -- one more variable turns the constant
one into zero. This would not be changed by complementing the output of
$f$: a nonempty cube has an even number of points, so
$\D_S(1+f)=\D_Sf$, and a hardness result for one target says nothing
about the other.

Secondly, complementing an input variable changes nothing at all,
because inside $S$ it merely reorders the summation and outside $S$ it
reorders the values $y$. We therefore complement a variable whenever this is
convenient, typically to make a variable that occurs only negatively
occur only positively.

Thirdly, when $f$ is a CNF, a certificate must contain a variable of
every clause that is not a tautology. Otherwise the clause can be
falsified using the outside variables alone. This makes $\D_Sf$ vanish
at that point. Note that a zero cube is under no such obligation.

Finally, we call a variable $x_i$ essential if $f$ depends on it. By
\cref{lem:anf}, this is the case exactly when $x_i$ lies in some support
of $f$, i.e., when $\{x_i\}$ is not a zero cube. In a positive unit-free
$2$CNF, every variable that occurs in a clause is essential, because
setting the other variable of that clause to $0$ and all remaining
variables to $1$ reduces the formula to $x_i$.

\subsection{Linear functions}
We use linear functions to convey a quick visualizable intuition for
the two problems. Since these may be analyzed by hand, they provide the
reader with an intuitive grasp of the problems. Let $f$ be the
indicator of the solutions of a consistent linear system $Hx=b$ over
$\F$. Fix a set $S$ of columns, and fix a value $y$ of the other
variables. The completions of $y$ to a solution of the system are the
solutions of $H_Sz=b+H_{\bar S}y$. A linear system has either no
solution or $2^{\,|S|-\operatorname{rank}H_S}$ of them, so every $y$ has
an odd number of completions exactly when the columns in $S$ are
linearly independent and span the remaining columns together with $b$.
Therefore when $S$ is a column basis of $H$, and only then,
$\D_Sf\equiv1$. Certificates are therefore the nonempty column bases, and
Gaussian elimination finds one.

Consider the case when the columns in $S$ are not independent of each
other. Flipping the variables of a dependent subset leaves $Hx$
unchanged. This pairs up the completions of every $y$, and all counts
are even. Had the columns been independent, the value $y$ obtained from
any solution would have had exactly one completion. Zero cubes are
therefore the dependent column sets, and the smallest one is the
minimum distance of the code $\ker H$, which is NP-hard to compute and
hard to approximate~\cite{bgr2025}.

\subsection{The approach}\label{sec:approach}
We begin by characterizing a certificate of a $2$CNF. We observe that
the fact that a certificate touches every clause implies that it is a
vertex cover of the graph whose edges are the clauses. A maximal
matching shows either (i) that no cover of size $k$ exists, or (ii) that
a set $T$ of at most $2k$ variables touches every clause.

Every other variable then necessarily interacts only with $T$, and in
one of $2^{O(k)}$ possible ways. Note also that variables that interact
in the same way are interchangeable. This reduces the search to
$2^{O(k^2)}$ candidate sets.

Further, a candidate $S$ can be checked without running through the
outside assignments: fixing $S$ leaves each clause with at most one
literal, so $\D_SF$ is a sum of $2^{O(k)}$ products of literals. Whether
such a sum is identically one is a question of linear algebra in
dimension $2^{O(k)}$. Note that a zero cube need not touch any clause,
and that the argument gives nothing for zero cubes.

For zero cubes, we establish hardness by translating a graph problem.
A multicolored clique~\cite{fhrv2009} is a clique in a graph whose
vertices are split into $r$ color classes. Every class contributes
exactly one vertex, and the clique has exactly $r$ vertices. In this
setting, the problem introduced by Fellows, Hermelin, Rosamond and
Vialette~\cite{fhrv2009} asks whether a given graph with $r$ color
classes contains a multicolored clique.

This problem has been shown to have no FPT algorithm unless
FPT$\,=\,$W[1]~\cite{fhrv2009,cygan2015}. We apply this framework to our
problem as follows. First, we turn such a graph into a positive $2$CNF
that has a zero cube of size $k$ when the graph has no multicolored
clique. Then, we ensure that when the graph has one, every zero cube of
the formula has more than $g(k)$ variables, for any computable $g$
fixed in advance. The hard part is to make picking a variable costly:
choosing a vertex of the graph should cost the zero cube many
variables, not one.

The usual device is to replace a variable by a product of $L$ new
ones. Although we would normally expect this to work, it charges
nothing. The terms that avoided the old variable avoid all $L$ new
ones, so picking one new variable is as good as picking all of them.
Instead, we need a block in which every proper subset is covered by
some term, while nothing covers the whole block. This way, only the
whole block is worth picking.

A triangle has this property: $uv+uw+vw$ covers every proper subset of
$\{u,v,w\}$. It is simultaneously the positive $2$CNF
$(u\vee v)(u\vee w)(v\vee w)$. Odd cliques do the same for every odd
size.

For the classical results we must convert a circuit, with some of its
inputs summed out, into a positive $2$CNF whose maximal terms are known
exactly. A parity-preserving encoding places the wanted terms among
those that contain all of the auxiliary variables. However, it also
creates other terms, any of which might be a small maximal term that was
never intended. We therefore use the crown construction to add a marker
variable for each auxiliary variable. The marker is joined to everything
except that variable. This plants a large term that misses exactly that
variable, so every unintended term sits inside a planted one and is no
longer maximal.

\subsection{Conventions}
Circuits use a fixed bounded-fan-in basis and list all their inputs. In
a gap statement the parameter is the supplied $k$ unless stated
otherwise. $\Sig$ is the second level of the polynomial hierarchy
relative to a parity oracle. Of the class $\Sig$ we need only one
complete problem. Given a circuit $R$,
decide whether
\begin{equation}\label{eq:source}
 \exists\alpha\in\{0,1\}^p\ \forall v\in\{0,1\}^q:
 \qquad h_\alpha(v):=\bigoplus_{w\in\{0,1\}^r}R(\alpha,v,w)=1 .
\end{equation}
\Cref{app:normal} shows that this problem is complete, although a $\Sig$
machine may query its oracle adaptively. Unused $\alpha$ or $v$ inputs
may be added so as to ensure $p,q\ge1$. An unused $w$ may not be added,
since it would make every $h_\alpha$ vanish.
\section{Finding a small certificate in a signed 2CNF}\label{sec:small-one}
This section proves the algorithmic half of \cref{thm:main-separation}.
Trying all sets of at most $k$ variables takes roughly $N^k$ time, which
is polynomial for each fixed $k$ but not fixed-parameter tractable. We
bring this down to $2^{O(k^2)}\poly(\ell)$.

\begin{theorem}\label{thm:small-one}
For a signed $2$CNF $F$, one can decide whether $\depth(F)\le k$, and
return a minimum certificate if there is one of size at most $k$, in
$2^{O(k^2)}\poly(\ell)$ time.
\end{theorem}

There are two obstacles, the number of candidate sets and the cost of
checking a single one. Both are removed by the observation that a
certificate touches every clause. Fixing the variables of a certificate
leaves each clause with at most one literal. This makes a candidate
easy to check, and the fact that the certificate is a vertex cover of
the clause graph ensures that candidates are few.

\paragraph{Checking one candidate.}
Fix a nonempty set $S$ with $s=|S|$, after removing tautological clauses
and repeated literals. If some clause contains no variable of $S$, then
$S$ is not a certificate. Otherwise consider the $2^s$ ways of fixing the
variables of $S$. Under each of them every clause that is not already
satisfied remains a single literal. Therefore, the cofactor is either $0$
or a product of literals in the remaining variables $y_1,\dots,y_t$. So
$\D_SF$ is a sum of at most $2^s$ such products. If $F$ expresses $x=y$
and $S=\{x\}$, for example, the two cofactors are $1+y$ and $y$, whose
sum is $1$, so $\{x\}$ is a certificate.

\begin{lemma}\label{lem:one-check}
Given a signed $2$CNF $F$ and a nonempty set $S$, one can decide whether
$\D_SF\equiv1$ in $2^{O(|S|)}\poly(\ell)$ time.
\end{lemma}
\begin{proof}
Adding the constant $1$, written as a product all of whose factors are
$1$, we must decide whether
\[
 G(y)=1+\D_SF(y)=\sum_{h=1}^{d}\prod_{j=1}^{t}q_{hj}(y_j)\equiv0,
 \qquad d\le2^s+1,\quad q_{hj}\in\{1,\ y_j,\ 1+y_j\}.
\]
Read the variables $y_1,y_2,\dots$ in order. After the first $j$ of them
have been assigned, each product is either alive, meaning that all of its
factors so far equal $1$, or dead, and we record this as a row vector
$\bigl(\prod_{i\le j}q_{hi}(y_i)\bigr)_{h}\in\F^d$. The number of
reachable rows may be exponential, but their span $L_j$ has dimension at
most $d$, and a basis of it suffices. If $M_j(b)$ denotes the diagonal
matrix with entries $q_{hj}(b)$, then
\[
 L_0=\spanf\{\mathbf1\},\qquad
 L_j=\spanf\bigl(L_{j-1}M_j(0)\cup L_{j-1}M_j(1)\bigr),
\]
since every reachable row extends by a $0$ or a $1$ and the maps
$M_j(b)$ are linear, so that $L_j$ is exactly the span of the rows
reachable after $j$ steps. At the end, $G(y)$ is the sum of the entries
of the final row, a linear functional, which vanishes on every reachable
row if and only if it vanishes on a basis of $L_t$. Building the products
costs $O(2^s\ell)$ and the linear algebra costs $O(td^3)$.
\end{proof}

The lemma is identity testing for a read-once oblivious branching
program of width $d$~\cite{razshpilka2005,forbesshpilka2013}.

\paragraph{Cutting down the candidates.}

\begin{proof}[Proof of \cref{thm:small-one}]
Reject if $k=0$, and remove tautologies, repeated literals and repeated
clauses. An empty clause means $F=0$, which has no certificate. Let $U$
be the set of variables occurring in unit clauses. Every certificate
contains $U$, so we reject if $|U|>k$.

Form the graph with a vertex for each variable and an edge for each
binary clause, disregarding signs, and greedily take a maximal matching
among the edges with both ends outside $U$. A certificate contains an
endpoint of every matching edge, so we reject if the matching has more
than $k-|U|$ edges. Otherwise let $T$ consist of $U$ and the endpoints of
the matching. Then $|T|\le2k$, and no clause has both of its variables
outside $T$, since such an edge could have been added to the matching.

A variable $v\notin T$ shares clauses only with variables of $T$. Its
type records, for each $t\in T$, which of the four clauses
$(t\vee v)$, $(t\vee\neg v)$, $(\neg t\vee v)$, $(\neg t\vee\neg v)$ are
present. There are at most $16^{2k}$ types, and exchanging two variables
of the same type leaves $F$ unchanged.

Now keep $T$ together with up to $k$ representatives of each type. This
pool contains at most $2k+k\cdot16^{2k}=2^{O(k)}$ variables, and any set
of at most $k$ variables can be moved into it, type by type, by
exchanges that do not affect whether the set is a certificate. The pool
has $2^{O(k^2)}$ nonempty subsets of size at most $k$. We test them in
order of increasing size with \cref{lem:one-check}, and the first success
is a minimum certificate.
\end{proof}

Keeping $k+2$ representatives of each type and deleting the rest with
their clauses yields a kernel of $2^{O(k)}$ bits (\cref{thm:kernel},
\cref{app:kernel}). Forbidden or mandatory variables are accommodated by
splitting the types and enlarging $U$. It is salient to point out that
the argument gives nothing for zero cubes, which need not touch any
clause. It also yields nothing for clause width three -- fixing a
certificate leaves a $2$CNF rather than a product of literals. We note
that this second limitation is inherent (\cref{thm:horn}).
\section{Small zero cubes are hard to find, even approximately}\label{sec:zero}
We now prove the hardness claimed in \cref{thm:main-separation}(b). To
do so, we begin with the observation that the algorithm of
\cref{sec:small-one} relies on the certificate touching every clause. A
zero cube however is under no such obligation. The following theorem
shows how much this costs.

\begin{theorem}\label{thm:zero-gap}
For every computable $g(k)\ge k$, distinguishing $\zeta(F)\le k$ from
$\zeta(F)>g(k)$ is coW[1]-hard with respect to $k$, even on positive
unit-free $2$CNF in which every variable is essential. The instances
produced by the reduction are satisfiable and come with a zero cube. The
reduction chooses an odd $L>g(k)$, and its two cases have optimum at most
$k$ and exactly $L$.
\end{theorem}

Recall the multicolored clique problem from \cref{sec:approach}. The
reduction we describe starts from its complement. We build a formula in
which summing over a particular set $A$ of $k$ variables gives the
indicator of the multicolored cliques. Therefore, $A$ is a zero cube of
size $k$ whenever there is no clique. When a clique exists we need every
zero cube to be large, which requires making the variables outside $A$
expensive.

\paragraph{Making a variable cost more than one.}
We begin by demonstrating how an obvious approach fails. To establish a
cost of 3 to pick variable $x$, the natural approach would be to replace
$x$ by a product $uvw$ of new variables. For certificates, this approach
is correct, because a maximal term contains a block either entirely or
not at all. For zero cubes, however, it fails for the following reason.

A term that did not contain $x$ contains none of $u,v,w$, so the set
$\{u\}$ is covered by exactly the terms that contained $x$. Picking $u$
alone has the same effect as picking the whole block. Observe that the
substitution left the proper subsets of the block uncovered. The remedy
is to make the terms that avoided $x$ cover every proper subset, while
still leaving the whole block uncovered. Put
\[
 P=uvw,\qquad E=uv+uw+vw .
\]
A term containing $x$ will contribute $P$, and a term not containing $x$
will contribute all three terms of $E$. Note that whereas every proper
subset of $\{u,v,w\}$ is contained in one of $uv,uw,vw$, the whole block
is contained in none of them. Hence, a zero cube that meets the block
must contain all of it.

Since $EP=P$, i.e., three copies of $uvw$ sum to one copy, the
substitution can be applied to all terms at once, which forms
$E\cdot f(P)$. Note that $E$ is the triangle
$(u\vee v)(u\vee w)(v\vee w)$, so the result is still a positive
$2$CNF.

The same works for every odd block size. Assign to each variable $x_i$ a
disjoint block $B_i$ of odd size $w_i\ge1$ and let
\[
 P_i=\prod_{b\in B_i}b,\qquad
 E_i=\sum_{b\in B_i}\ \prod_{c\in B_i\setminus\{b\}}c,
 \qquad
 f^w=\Bigl(\prod_iE_i\Bigr)\,f(P_1,\dots,P_n).
\]
The polynomial $E_i$ equals $1$ exactly when at most one variable of
$B_i$ is $0$. Hence, it is the clique formula $\bigwedge_{b\ne c}(b\vee c)$
on $B_i$, and $E_i=1$ when the block is a singleton.

\begin{lemma}\label{lem:zero-block}
For $I\subseteq\bigcup_iB_i$ let $J(I)=\{i:B_i\subseteq I\}$ be the set of
blocks contained in $I$. Then, with the convention $\D_\varnothing f=f$,
\[
 \D_If^w\equiv0\iff\D_{J(I)}f\equiv0 .
\]
If $f\ne0$, then
$\zeta(f^w)=\min\{\sum_{i\in J}w_i:\ J\ne\varnothing,\ \D_Jf\equiv0\}$,
and if $f$ is a positive unit-free $2$CNF, then so is $f^w$.
\end{lemma}
\begin{proof}
Since $w_i$ is odd, $E_iP_i=w_iP_i=P_i$, and therefore
$f^w=\sum_{A:\,c_A=1}\prod_{i\in A}P_i\prod_{i\notin A}E_i$. Expanding
the right-hand side produces one term for each choice of a support $A$
and, for each $i\notin A$, of an omitted variable in $B_i$. Different
choices produce different supports, since $A$ can be read off as the set
of blocks that appear entirely and the omitted variable can be read off
in each remaining block, so no cancellation occurs.

Fix a support $A$ of $f$. Some term produced by $A$ contains $I$ if and
only if $A\supseteq J(I)$: a block contained in $I$ must appear entirely
in the term, which forces its index into $A$, and for $i\notin A$ the
block $B_i$ is not contained in $I$, so the omitted variable can be
chosen outside $I$. Hence some support of $f^w$ contains $I$ if and only
if some support of $f$ contains $J(I)$, and \cref{lem:anf} gives the
first claim.

For the second claim let $f\ne0$ and let $I$ be a zero cube of $f^w$.
Then $J(I)\ne\varnothing$, because $\D_\varnothing f=f\not\equiv0$.
Removing the partial blocks from $I$ leaves $J(I)$ unchanged and
shrinks $I$, and conversely every zero cube $J$ of $f$ lifts to the
zero cube $\bigcup_{i\in J}B_i$ of $f^w$. This gives the formula for
$\zeta(f^w)$. For the last claim, $E_i$ is the clique on $B_i$, and a
clause $x_i\vee x_j$ of $f$ becomes
$P_i\vee P_j=\bigwedge_{b\in B_i,\,c\in B_j}(b\vee c)$, so that blocks
become cliques and edges become complete bipartite graphs.
\end{proof}

\paragraph{The clique formula.}
\begin{proof}[Proof of \cref{thm:zero-gap}]
Let the Multicolored Clique instance have nonempty color classes
$V_1,\dots,V_r$ with $r\ge2$, and set $k=r$. Two distinct vertices are
incompatible if they belong to the same class, or to different
classes without being adjacent. Introduce a variable $y_v$ for each
vertex, with the meaning that $v$ is in the clique, and a
selector $a_i$ for each class, and let
\[
 Q(a,y)=
 \bigwedge_{\{u,v\}\text{ incompatible}}(\neg y_u\vee\neg y_v)
 \ \wedge\ \bigwedge_{i}\ \bigwedge_{v\in V_i}(a_i\vee\neg y_v).
\]
The first group of clauses requires the chosen vertices to be pairwise
compatible. The second group forces $a_i=1$ as soon as some vertex of
class $i$ is chosen and leaves $a_i$ unconstrained otherwise, so that for
fixed $y$ the sum over $a_i$ is $1$ in the first case and $0$ in the
second. With $A=\{a_1,\dots,a_r\}$ this gives
\begin{equation}\label{eq:zero-clique}
 \D_AQ(y)=
 \bigwedge_{\{u,v\}\text{ incompatible}}(\neg y_u\vee\neg y_v)
 \cdot\prod_i\Bigl(\bigvee_{v\in V_i}y_v\Bigr)=:C_G(y),
\end{equation}
the indicator of the multicolored cliques.

Complementing every $y_v$ turns $Q$ into a positive $2$CNF $Q^+$ and, by
\cref{sec:basic}, changes no derivative. Add an unused variable $d$ and
apply \cref{lem:zero-block} with weight $1$ on each selector and an odd
weight $L\ge3$, $L>g(r)$, on every vertex variable and on $d$. Call the
result $F$.

If there is no clique, then $C_G\equiv0$, hence $\D_AQ^+\equiv0$, and
since the selectors form singleton blocks, $A$ is a zero cube of $F$ of
size $r=k$.

If there is a clique, let $I$ be any set of fewer than $L$ variables.
Such a set contains neither a full vertex block nor the full block of
$d$, so $J(I)\subseteq A$. Were $I$ a zero cube, the lemma would give
$\D_{J(I)}Q^+\equiv0$ with $J(I)\ne\varnothing$ (as $Q^+\ne0$), and then
$\D_AQ^+=\D_{A\setminus J(I)}\D_{J(I)}Q^+\equiv0$, contradicting
$C_G\not\equiv0$. Hence $\zeta(F)\ge L$. On the other hand the block of
$d$ is a zero cube, because $Q^+$ does not depend on $d$, so
$\zeta(F)=L$, and this block is the zero cube supplied with the instance.

The all-ones assignment satisfies $F$. No vertex of the graph of $F$ is
isolated, since each block of size $L\ge3$ is a clique and each selector
is adjacent to the block of some vertex of its nonempty class, so every
variable is essential by \cref{sec:basic}. With $v=\sum_i|V_i|$ the
formula has $r+(v+1)L$ variables and $O(v^2L^2)$ clauses, and $L$ depends
only on $r$, so this is an FPT reduction.
\end{proof}

Notice a cogent detail here: the supplied zero cube has size $L$ in both
cases, and the optimum is at most $L$ in both cases. Therefore, an
algorithm whose running time depends on the optimum rather than on $k$
gains nothing.

\begin{corollary}\label{cor:zero-approx}
Unless $\mathrm{FPT}=\mathrm{W[1]}$, no algorithm always returns a zero
cube of size at most $\rho(\zeta(F))\zeta(F)$ in time
$h(\zeta(F))\ell^{O(1)}$ on positive $2$CNF with a zero cube, for any
computable $h$ and $\rho\ge1$. Unless P=NP, there is no polynomial-time
$N^{1-\varepsilon}$ approximation for any fixed $0<\varepsilon<1$, and no
$\ell^{1/2-\varepsilon}$ approximation for any fixed $0<\varepsilon<1/2$.
\end{corollary}

The proof, given in \cref{app:zero-approx}, chooses $L$ so large that an
approximate answer in the small case must have fewer than $L$ variables.
This is impossible in the large case. The exponent $1/2$ reflects the
$L^2$ clauses of a block. The $3$CNF construction of \cref{app:zero} uses
no blocks and reaches $\ell^{1-\varepsilon}$.
\section{Lifting a circuit into positive 2CNF}\label{sec:lift}
Of the three main results we present, the last concerns classical
complexity: BCD is $\Sig$-complete on positive unit-free $2$CNF and hard
to approximate. Since hardness has to start from the circuit
problem~\eqref{eq:source} and end in positive $2$CNF, a
parity-preserving encoding of circuits is required. The difficulty is
that BCD concerns maximal terms. An encoding that produces the terms we
want also produces others, and any of them may be a small maximal term
that was never intended.

Hence we proceed in three steps. A compiler makes the wanted terms
exactly those that contain all of the auxiliary variables $M$. A crown
plants a larger term above every unwanted one, and weights give the
original variables a price. The outcome, \cref{thm:lift}, lists the
certificates of the final formula completely.

\paragraph{The compiler.}
Let $C(X,Z)$ be a circuit and let $F(X)=\bigoplus_zC(X,z)$ be the function
obtained by summing out the inputs $Z$.

\begin{lemma}[Positive parity compiler]\label{lem:compiler}
There is a positive unit-free $2$CNF $Q(X,M)$ with
\begin{equation}\label{eq:compiler}
 \D_MQ=F,
\end{equation}
in which every clause has two distinct variables, at least one of them in
$M$. One may take $m=|M|$ odd and at least $3$. The number of auxiliary
variables and the number of clauses are $O(s)$, where $s$ is the size of
$C$ including its declared inputs.
\end{lemma}

The proof is in \cref{app:compiler}: gate variables are introduced in the
usual way, each clause is replaced by binary clauses on a private
variable whose parity reproduces it, and negative literals are
eliminated by Faben's triangle~\cite{faben2008}. By \cref{lem:anf},
equation~\eqref{eq:compiler} says that the terms of $Q$ containing all of
$M$ are the terms $x_Mx_A$ with $x_A$ a term of $F$. The remaining terms
of $Q$ each miss some variable of $M$, and we have no control over them.

\paragraph{The crown.}
Write $M=\{\tau_1,\dots,\tau_m\}$, add an unused variable $d$, and let
$Y=X\cup M\cup\{d\}$. For each $\tau_i$ introduce a marker $t_i$,
joined by a positive clause to every variable of $Y$ other than
$\tau_i$:
\begin{equation}\label{eq:crown}
 H=Q\ \wedge\ \bigwedge_{i=1}^m\ \bigwedge_{y\in Y\setminus\{\tau_i\}}(y\vee t_i).
\end{equation}
Setting $t_i=0$ forces every variable of $Y$ except $\tau_i$ to $1$,
which satisfies $Q$ regardless of $\tau_i$ because each clause of $Q$ has
two distinct variables. Hence $H$ has a term consisting of all of
$Y$ but $\tau_i$ together with all markers but $t_i$, and every unwanted
term of $Q$, which misses some $\tau_i$, is contained in the
corresponding planted term. The variable $d$ serves to keep the planted
terms distinct from anything that $Q$ already had. Writing
$T=\{t_1,\dots,t_m\}$, the maximal supports of $H$ are as follows.

\begin{lemma}[The crown's maximal supports]\label{lem:crown}
Let $N=|Y|+m$. The certificates of $H$ are exactly
\begin{align}
 &T\cup S,&&S\in\Max(Q),\ M\subseteq S,\label{eq:corefamily}\\
 &(Y\cup T)\setminus\{\tau_i\},&&i\in[m],\label{eq:oldguards}\\
 &(Y\cup T)\setminus\{t_i\},&&i\in[m].\label{eq:markerguards}
\end{align}
The sets in \eqref{eq:oldguards} and \eqref{eq:markerguards} are $2m$
guards, each of size $N-1$. Every variable of $H$ is essential.
\end{lemma}

The first family consists of the wanted terms with all markers attached,
and such a term is maximal in $H$ exactly when it was maximal in $Q$. The
guards are the planted terms and their mirror images. The full term
$x_{Y\cup T}$ is absent because $m$ is odd, which is why the guards are
maximal. The calculation is in \cref{app:crown-proof}.

\paragraph{Weights.}
Our reduction needs picking some variables to cost more than others.
The proof of \cref{thm:hardness} gives weight $1$ to the variables that
encode the existential bits $\alpha$, and weight $L$ to each universal
variable. Then every certificate of a NO instance is larger by at least
$L$ than the smallest certificate of a YES instance. Taking $L$ large
turns this difference into an approximation gap.

We give a variable $x\in X$ the weight $w_x$ as follows. We replace $x$
in $Q$ by the product of a block $B_x$ of $w_x$ new variables. Since the
blocks are disjoint and new, different terms of $Q$ stay different after
the substitution, so no two of them cancel. Also, one term contains
another after the substitution exactly when it did before. Therefore,
picking $x$ costs exactly $w_x$ for certificates.

Hence, a maximal term of the substituted formula contains each block
either entirely or not at all. Zero cubes do not behave this way, which
is why \cref{sec:zero} needed odd cliques instead.

We carry out this substitution before adding the crown, so that the
markers are joined to the block variables as well. The result is still
a positive $2$CNF. Each clause of $Q$ contains at most one variable of
$X$, and a clause $(\prod_{b\in B_x}b)\vee y$ splits into the binary
clauses $\bigwedge_{b\in B_x}(b\vee y)$. The auxiliary variables $M$
are untouched.

\begin{theorem}[Weighted maximal-support lifting]\label{thm:lift}
Given a circuit $C(X,Z)$ and positive integer weights $w_x$ in unary, one
can construct in polynomial time a positive unit-free $2$CNF $H$ whose
certificates are exactly
\[
 M\cup T\cup\bigcup_{x\in A}B_x\quad(A\in\Max(F)),
 \qquad\text{together with the $2m$ guards},
\]
where $F=\bigoplus_zC$, $|B_x|=w_x$, $|M|=|T|=m=O(s)$, and the guards are
those of \eqref{eq:oldguards}--\eqref{eq:markerguards} for
$Y=M\cup\{d\}\cup\bigcup_xB_x$. In particular $N=2m+\sum_xw_x+1$ and
\[
 \depth(H)=\min\Bigl\{2m+\min_{A\in\Max(F)}\sum_{x\in A}w_x,\ N-1\Bigr\},
\]
an empty minimum being read as $+\infty$. If every $w_x\le L$, then $H$ has
$O(sL)$ variables and $O(s^2L)$ clauses. All variables are essential, and
the guards are explicit certificates of size $N-1$.
\end{theorem}
\begin{proof}
After the substitution, the terms of the compiler that contain $M$ are
$M\cup\bigcup_{x\in A}B_x$ for the supports $A$ of $F$, and such a term
is contained only in terms that also contain $M$, so it is maximal
exactly when $A$ is maximal in $F$. \Cref{lem:crown} then lists the
certificates of $H$ as claimed. The substituted compiler has $O(sL)$
clauses, and the crown adds $m(|Y|-1)=O(s^2L)$ more.
\end{proof}

\paragraph{The reduction.}
\begin{theorem}\label{thm:hardness}
BCD is $\Sig$-complete on positive unit-free $2$CNF. For every fixed
$0<\varepsilon<1$, a polynomial-time algorithm that always returns a
certificate of size at most $N^{1-\varepsilon}\depth(H)$, or of size at
most $\ell^{1-\varepsilon}\depth(H)$, would imply $\Sig=\mathrm P$. The
hard instances have every variable essential and an explicit certificate
of size $N-1$.
\end{theorem}
\begin{proof}
Take an instance~\eqref{eq:source}. We arrange for the choice of a
certificate to amount to the choice of $\alpha$, representing each bit
$\alpha_i$ by a pair of variables $a_i,b_i$, where choosing $a_i$ means
$\alpha_i=1$ and choosing $b_i$ means $\alpha_i=0$. Let
\[
 C(u,w,a,b,v)=R(u,v,w)\prod_{i=1}^p\bigl[(u_i\wedge a_i)\vee(\neg u_i\wedge b_i)\bigr]
\]
and sum out $u$ and $w$. For $u=\alpha$ the product is the single term
$x_{A_\alpha}$ that takes the chosen member of each pair, so that
\begin{equation}\label{eq:selector}
 F(a,b,v)=\sum_{\alpha\in\{0,1\}^p}x_{A_\alpha}\,h_\alpha(v).
\end{equation}
Two different values of $\alpha$ disagree in some pair, so a term arising
from one can never contain a term arising from the other, and the maximal
supports arising from $\alpha$ are the sets $A_\alpha\cup J$ with
$J\in\Max(h_\alpha)$.

Apply \cref{thm:lift} with weight $1$ on every $a_i,b_i$ and weight $L$ on
every universal variable $v_j$, and put $K=2m+p$. A certificate arising
from $A_\alpha\cup J$ costs $K+L|J|$. If some $h_\alpha$ is identically
one, its only maximal support is $J=\varnothing$ and the cost is exactly
$K$. If no $h_\alpha$ is identically one, then every nonzero $h_\alpha$
is nonconstant, all of its maximal supports are nonempty, and every
certificate of this kind costs at least $K+L$. The guards cost
$N-1=K+p+qL\ge K+L$. A YES instance thus has $\depth(H)=K$ and a NO
instance has $\depth(H)\ge K+L$. With $L=1$ this is a polynomial-time
reduction and proves hardness. Membership in $\Sig$ is immediate: guess
the set, universally choose the outside assignment, and ask a parity
oracle (\cref{app:normal} gives the fixed-length encoding). For the approximation statements take $L$ to be a
sufficiently large polynomial in $s$. An approximation within either
ratio must then return a certificate of size below $L$ on a YES instance
and cannot do so on a NO instance (\cref{app:one-approximation}).
\end{proof}

The same construction gives $\Sig$-completeness for zero cubes
(\cref{thm:zero-classical}, \cref{app:zero}). There the guards force a
zero cube to contain all of $M\cup T$, and the rest of the argument goes
through unchanged.
\section{Other representations}\label{sec:other}
The results so far concern CNF formulas, and it is natural to ask how
much they depend on that choice. To examine how far they generalize, we
consider several other scenarios, such as binary constraints rather than
clauses, and functions given by rounds of quadratic maps in the
cryptographic setting. We also consider two structural restrictions,
treewidth and branching-program width, under which the problem admits
algorithms. Length restrictions prevent us from listing all of our
proofs here. We encourage readers to refer to our appendix for
full details.

\paragraph{Binary constraint languages.}
Our hardness results were proved for the single binary constraint OR,
that is, for positive $2$CNF. A relation of arity $a$ is a set of
tuples in $\{0,1\}^a$. For example, OR is the set $\{01,10,11\}$. A
constraint applies a relation to a tuple of variables, and is satisfied
when their values form one of its tuples. A Boolean constraint language
is a finite set $\Gamma$ of relations. An instance over $\Gamma$ is a
conjunction of constraints whose relations come from $\Gamma$. A
positive $2$CNF is thus an instance over $\{\mathrm{OR}\}$.

A relation is affine if its tuples are exactly the solutions of a system
of linear equations over $\F$. For example, $x=y$ and $x\ne y$ are
affine. Among relations of arity at most two, only those with exactly
three tuples fail to be affine. There are four of them, one for each
excluded tuple. OR excludes $00$, NAND excludes $11$, the implication
$x\to y$ excludes $10$, and its reverse excludes $01$.

\begin{theorem}\label{thm:binary}
For a finite Boolean constraint language $\Gamma$ of arity at most two,
BCD$(\Gamma)$ is in P if every relation in $\Gamma$ is affine, and
$\Sig$-complete otherwise. In every case there is a $2^{O(k^2)}\poly(\ell)$
algorithm with respect to the certificate size $k$.
\end{theorem}

The affine case we consider is the example of \cref{sec:overview}.
Recall that in this scenario, certificates are the nonempty column
bases, and Gaussian elimination finds one. OR is \cref{thm:hardness}.
NAND reduces to OR by complementing every variable, which changes no
derivative.

Implication needs more work. The relation $x\to y$ is the clause
$\neg x\vee y$, so it treats its two variables differently. Complementing
$x$ turns it into OR. However, a variable can be the tail of one
implication and the head of another, so in general no single
complementation turns every implication into OR. The crown also depends
on how clauses are satisfied. An OR clause is satisfied by setting
either of its variables to $1$. An implication is satisfied by setting
its tail to $0$ or its head to $1$. Hence the crown has to be rebuilt to
account for both moves. \Cref{app:implication} carries this out. It
proves the lifting theorem with implications only, even when the
implication graph is acyclic and has height at most four.

Finally, the parameterized algorithm of \cref{thm:small-one} applies to
every binary language, because any binary relation is a signed $2$CNF of
constant size.

\paragraph{Quadratic rounds.}
In cryptographic applications, the function $f$ is typically one output
bit of a cipher, viewed as a function of its key and public input bits.
It is given as a sequence of rounds. We take a round to be an explicitly
listed map over $\F$ all of whose coordinates have degree at most two.
The dimensions may change from round to round, the whole description
counts towards $\ell$, and $n$ denotes the number of initial inputs.
Since a round can at most double the degree of each coordinate, as a
polynomial in the initial inputs, after $r$ rounds the function has
degree at most $2^r$. Hence one can simply compute all of its
$n^{O(2^r)}$ coefficients. Unless $\mathrm{NP}\subseteq\mathrm{SUBEXP}$,
the exponent cannot be made subexponential in $r$.

\begin{theorem}\label{thm:rounds}
BCD on $r$ quadratic rounds is solvable in $n^{O(2^r)}\poly(\ell)$ time.
For every fixed $c>0$, unless $\mathrm{NP}\subseteq\mathrm{SUBEXP}$, no
algorithm runs in time $n^{a(r)}\ell^{O(1)}$ with $a(r)=2^{o(r)}$, even on
instances with
$2^r=\Theta\bigl(n^{1/(c+1)}/(\log n)^{c/(c+1)}\bigr)$.
\end{theorem}

We prove the lower bound in \cref{app:rounds} by a reduction from
$3$SAT on $N$ variables. We do so by splitting the $N$ variables into
groups of about $c\log N$ variables each, and introducing one input for
every group and every assignment to it. Picking one input from each
group then describes a complete assignment. Note that this only needs
$N^{O(c)}$ inputs and $O(\log N)$ rounds. An algorithm whose exponent is
only $2^{o(r)}$ would therefore solve $3$SAT in time $2^{N^{o(1)}}$.

\paragraph{Treewidth.}
A large formula may nevertheless decompose into pieces that share few
variables. The treewidth of the formula directly measures how far this
is the case. In comparison to most standard problems on tree
decompositions, our case is complicated by the fact that the cube is
chosen once and must work for every outside assignment.

\begin{theorem}\label{thm:width}
Let $f$ be a product of local Boolean factors, given with a tree
decomposition of width $w$ of its primal graph, in which two variables are
adjacent if they occur in a common factor. A minimum certificate can be
found in $2^{2^{O(w)}}\poly(\ell)$ time. Variables may be declared
forbidden, or summed out at zero cost, so the same bound holds for
circuits with respect to the width of their gate-constraint graph. Unless
$\mathrm{NP}\subseteq\mathrm{SUBEXP}$, BCD on circuits admits no
$2^{2^{o(w)}}\poly(\ell)$ algorithm, even when a path decomposition of
width $w$ is supplied.
\end{theorem}

Note, therefore, that our algorithm is a dynamic program over a tree
decomposition, whose bags hold at most $w+1$ variables each. To see what
the dynamic program must remember, consider the part of the formula
below a bag $B$, and fix the part of the cube that lies there. Each
assignment to the outside variables below $B$ then leaves a table of
$2^{|B|}$ bits, one for each assignment to $B$.

Since the dynamic program cannot remember these entries separately
(because it would lose track of which entries came from the same outside
assignment), it must remember the whole family of tables, which can be
very large. Instead, we store the smallest affine subspace containing
the family, called its affine hull.

Our approach works because the dynamic program never needs the family
itself, only its affine hull. To see this, recall what the dynamic
program does at each step. It either sums out a chosen variable, keeps
both restrictions of an unchosen variable, or joins two children.
Summing out is a linear map on tables. Keeping both restrictions
produces the union of two linear images of the family. Joining two
children multiplies their tables pointwise, which is a bilinear map. For
all three operations, the affine hull of the result is determined by
the affine hulls of the inputs. Hence the dynamic program can pass hulls
from the leaves to the root without ever storing a family.

At the root, we may assume the bag is empty, so each table is a single
bit. The cube is a certificate exactly when every outside assignment
yields $1$, that is, when the hull is $\{1\}$. It is a zero cube exactly
when the hull is $\{0\}$. This is why we use affine hulls rather than
linear spans. The span of $\{1\}$ equals the span of $\{0,1\}$, so a
span could not tell a certificate from a cube that fails. Finally, the
number of possible hulls at a bag is the number of affine subspaces of
$\F^{2^{w+1}}$, which is doubly exponential in $w$. This bounds the
number of states, and gives the running time
$2^{2^{O(w)}}\poly(\ell)$.

For the lower bound, we build a circuit of pathwidth $O(\log n)$. It
reads an address naming one clause of a SAT instance, and checks only
that clause against a proposed assignment. The certificate condition
quantifies over all outside assignments, and hence over all addresses.
This supplies the quantification over all clauses. The details are in
\cref{app:width}. The correct dependence on treewidth for $2$CNF remains
open.

For ordered branching programs, the fixed variable order allows a
cheaper version of the same dynamic program. For each candidate set, it
keeps the span of the rows of its strict supersets. This gives an
algorithm running in time $2^{O(w^2)}n$ in the width $w$
(\cref{thm:branching}, \cref{app:branching}). Unless
$\mathrm{NP}\subseteq\mathrm{SUBEXP}$, no algorithm runs in time
$2^{w^{o(1)}}$ times a polynomial (\cref{prop:obdd-hard}).

\clearpage
\appendix
\begin{center}{\Large\bfseries Appendix}\end{center}
\noindent
In this appendix, we present the proofs that were deferred from the main
text. For the convenience of the reader, the table below lists what each
part contains.

\begin{center}
\small
\begin{tabular}{@{}ll@{}}
\toprule
Appendix & What it proves\\
\midrule
A & the positive parity compiler and the crown lemma; size calculation for approximation\\
B & the quantified-parity normal form behind $\Sig$, and membership\\
C & zero cubes: the approximation corollary, classical completeness, fixed thresholds, $3$CNF\\
D & the $2^{O(k)}$-bit kernel for signed $2$CNF\\
E & the treewidth algorithm and the lower bound for circuits\\
F & the clause-width boundary: Horn $3$CNF, fixed thresholds, checking a given set\\
G & the implication version of the lifting theorem and the binary-language dichotomy\\
H & the number of quadratic rounds\\
I & ordered branching programs\\
J & hardness with a strongly collapsible independence complex\\
\bottomrule
\end{tabular}
\end{center}

\section{Details of the lifting construction}\label{app:compiler}
The lifting theorem of \cref{sec:lift} rests on two lemmas, and in this
appendix we give their proofs. We begin with the compiler, which turns a
circuit into a positive $2$CNF, and then prove the crown lemma, which
lists the maximal supports of the final formula. Lastly, we carry out the
size calculation that turns the reduction of \cref{thm:hardness} into an
approximation gap.

\subsection{Converting a circuit to positive binary clauses}
We now prove \cref{lem:compiler}. The proof converts the circuit in three
passes, and none of the passes touches the free inputs $X$. Firstly, we
express the evaluation of the circuit by clauses. Secondly, we replace
each clause by binary clauses whose auxiliary assignments have the
correct parity. Thirdly, we remove the negative literals.

\paragraph{Circuit constraints.}
We introduce a variable for every gate, impose its truth table by a
constant number of CNF clauses, and pin the output to one. Note that
once the inputs are fixed, the value of every gate is determined. Hence
there is exactly one satisfying assignment of the gate variables when the
circuit outputs one, and none when it outputs zero. Summing over the gate
variables therefore recovers the output of the circuit.

\paragraph{Replacing a clause by binary clauses.}
Let $B=\ell_1\vee\cdots\vee\ell_t$ be one of these clauses. To replace it
by binary clauses without disturbing any parity, we introduce a variable
$e_B$ that is private to this clause, and replace $B$ by
\[
 P_B=\bigwedge_{j=1}^t(\neg\ell_j\vee e_B).
\]
Consider the two cases. If $B$ is false, every $\ell_j$ is zero and $e_B$
may take either value, so the two values cancel modulo two. If $B$ is
true, some $\ell_j$ is one and forces $e_B=1$, which leaves exactly one
allowed value. Consequently, $\bigoplus_{e_B}P_B=B$. Since different
clauses have different private variables, their parity sums can be taken
independently of each other. Note that this also turns unit clauses into
binary clauses -- constants in the circuit are represented by pinned
gates, so we never need an empty clause.

\paragraph{Removing negative literals.}
For every variable $t$ of the binary formula, we add two new variables
$\bar t$ and $\eta_t$ together with the positive triangle
\[
 (t\vee\bar t)(t\vee\eta_t)(\bar t\vee\eta_t).
\]
We then replace every occurrence of $\neg t$ by $\bar t$. Observe that at
this stage $\bar t$ is merely a new name, and nothing in the clauses
forces it to be the complement of $t$. The triangle enforces the
complement through parity instead: summing it over $\eta_t$ gives
$t+\bar t$, which is one exactly when $\bar t=1+t$. Summing over $\bar t$
thus restores the negative literal in the rest of the formula. This is
Faben's triangle construction~\cite{faben2008}.

We let $M$ contain every variable except those of $X$, i.e., the
original inputs $Z$, the gate variables, the private variables $e_B$, and
the two variables added by each triangle. Summing first over the triangle
variables restores the signed binary formula. Summing over the variables
$e_B$ then restores the circuit constraints, and summing over the gate
variables and $Z$ finally gives $F(X)$. This proves $\D_MQ=F$.

Further, every clause has two distinct variables, and every clause meets
$M$. A clause inherited from $P_B$ contains $e_B$ or its new complement
variable, whereas a triangle clause contains one of the new triangle
variables. Since the circuit basis has bounded fan-in, each of the three
passes increases the size by at most a constant factor.

The crown also needs $|M|$ to be odd and at least three, which we can
always arrange. If $|M|$ is even, we add a disjoint path on three new
vertices and put its variables in $M$. Its positive edge formula has five
satisfying assignments, so summing over these three variables multiplies
the parity by one, and $|M|$ becomes odd. If $|M|=1$, we instead add a
disjoint edge, which has three satisfying assignments and brings $|M|$ up
to three. This completes the proof of \cref{lem:compiler}.

\subsection{The maximal supports of the crown}\label{app:crown-proof}
We now prove \cref{lem:crown}. Recall that $Q$ is positive, that its
clauses are binary on distinct variables, and that it does not use the
dummy variable $d$. For the construction~\eqref{eq:crown} we write
\[
 P=x_Y,\qquad P_i=x_{Y\setminus\{\tau_i\}},\qquad
 t_T=\prod_{i=1}^m t_i,\qquad t_{-i}=\prod_{j\ne i}t_j.
\]
We determine $H$ by fixing its markers. If all markers are one, the added
clauses impose no condition, and we are left with $Q$. If only $t_i$ is
zero, the added clauses force every old variable except $\tau_i$ to one.
Notice that this assignment satisfies $Q$ regardless of $\tau_i$, because
every clause of $Q$ has a second variable, and that variable is set to
one. What remains is $P_i$. If two or more markers are zero, their
clauses together force every old variable to one, which leaves $P$. We
can therefore record the three cases as follows:
\[
 H=\begin{cases}
 Q&\text{if every marker is one},\\
 P_i&\text{if $t_i$ is the only marker at zero},\\
 P&\text{if at least two markers are zero}.
 \end{cases}
\]
The indicator that all markers are one is $t_T$, and the indicator that
only $t_i$ is zero is $(1+t_i)t_{-i}$. The remaining assignments to the
markers are indicated by $1+t_T+\sum_i(1+t_i)t_{-i}$. Hence
\[
 H=Qt_T+\sum_iP_i(1+t_i)t_{-i}
       +P\left(1+t_T+\sum_i(1+t_i)t_{-i}\right).
\]
Expanding this expression, and using the fact that $m$ is odd, gives
\begin{equation}\label{eq:crownanf}
 H=P+\sum_i(P_i+P)t_{-i}
       +t_T\left(Q+\sum_iP_i\right).
\end{equation}
In particular, the full monomial $Pt_T$ cancels, since its coefficient
before cancellation is $m+1$.

Now consider the two kinds of near-full monomials, $t_TP_i$ and
$Pt_{-i}$. Each of them has size $N-1$, and they are all distinct.
Neither kind can cancel with a term of $t_TQ$: the first contains $d$,
which $Q$ does not use, whereas the second omits a marker, and every term
of $t_TQ$ contains all of the markers. Since there is no full monomial
above them, they are maximal. These are exactly the guards in
\eqref{eq:oldguards} and \eqref{eq:markerguards}. Note that the other
explicit terms, $P$ and $P_it_{-i}$, are contained in guards.

We are left with the terms of $t_TQ$. If such a term misses some
$\tau_i$, it is strictly contained in the guard $t_TP_i$, which contains
its variables and also $d$. If it contains all of $M$, it cannot lie
inside any guard, because each guard misses one variable of $M\cup T$.
The only terms that can contain it are therefore other terms of $t_TQ$.
Hence it is maximal in $H$ exactly when its old support is maximal in
$Q$. This proves \eqref{eq:corefamily}, together with the completeness of
the list.

Finally, every variable occurs in at least one guard, and therefore in
the unique multilinear polynomial of $H$. Changing such a variable
changes the function on some input, so every variable is essential.

\subsection{The size calculation for approximation hardness}\label{app:one-approximation}
The reduction in the proof of \cref{thm:hardness} has optimum $K$ in the
YES case and at least $K+L$ in the NO case. Our idea is to choose $L$ so
large that an approximation must still return a set with fewer than $L$
variables on a YES instance. The size of the returned set then separates
the two cases.

Let $s\ge2$ be the size of the unweighted source. For an explicit binary
encoding of the clause list, the construction gives
\[
 K=O(s),\qquad N=O(sL),\qquad
 \ell=O(s^2L\log(sL)).
\]
Fix $0<\varepsilon<1$, choose an integer
$A\ge\lceil4/\varepsilon\rceil$, and set $L=s^A$. Then
\[
 \frac{K\ell^{1-\varepsilon}}{L}
 =O_\varepsilon\left(s^{3-2\varepsilon-A\varepsilon}
                     (\log s)^{1-\varepsilon}\right)
 \longrightarrow0.
\]
The same conclusion holds with $N^{1-\varepsilon}$ in place of
$\ell^{1-\varepsilon}$, since $N=O(sL)$ gives an even smaller bound.
Therefore, for all sufficiently large source instances, an approximation
with either ratio must return a certificate with fewer than $L$ variables
in the YES case, whereas no such certificate exists in the NO case. The
finitely many smaller source sizes can be solved directly.

Since $A$ is fixed, the reduction has polynomial size. An algorithm with
either ratio would therefore decide a $\Sig$-complete problem in
polynomial time, which proves the approximation part of
\cref{thm:hardness}.

\section{The quantified-parity normal form}\label{app:normal}
Our classical completeness results reduce from the quantified circuit
problem~\eqref{eq:source}. In this appendix, we explain why that problem
is complete for $\Sig$, and why BCD belongs to $\Sig$. The only subtlety
is that a $\Sig$ machine may query its parity oracle adaptively, i.e., a
later query may depend on the answers to earlier ones.

\subsection{Combining adaptive parity queries}
The closure $\mathrm P^{\oplus\mathrm P}=\oplus\mathrm P$ is
standard~\cite{beigel1992}. We nevertheless include the transcript
argument here, because it shows directly why adaptivity causes no
difficulty.

Consider a deterministic polynomial time machine with a parity oracle,
padded so that it always makes the same polynomial number of queries.
Suppose that we predict all of the oracle answers in advance, and write
them as a bit string $b$. The machine can then be simulated without its
oracle. Let $a_b$ be the acceptance bit of this simulation, and let
$q_i(b_{<i})$ denote its $i$th query. Note that this query may depend on
the earlier predicted answers. If $\chi(q)$ denotes the true answer to a
query $q$, then
\[
 1+b_i+\chi(q_i(b_{<i}))
\]
is one exactly when the $i$th prediction is correct. The product of all
these factors is therefore one for the single transcript that is correct
throughout, and zero for every other one. Consequently, the true
acceptance bit of the machine is
\[
 \bigoplus_b a_b\prod_i
       \bigl(1+b_i+\chi(q_i(b_{<i}))\bigr).
\]

We observe that this expression is itself a parity predicate. Adding the
constant one to a factor can be implemented by adding one accepting
branch. A product of parities is implemented by choosing independent
witnesses for its factors, and accepting only when all of their
predicates accept. The outer parity sum then chooses $b$. Since there are
polynomially many factors, each with a witness of polynomial length, the
resulting witness length is polynomial. It is salient to point out that
when fixed witness lengths require padding, the extra positions have to
be forced to zero. Leaving them free would duplicate every witness and
destroy its parity.

\subsection{The complete problem and membership}
The standard relativized characterization of the second level gives an
existential block, followed by a universal block and a deterministic
computation with a parity oracle. The closure we just proved replaces the
final computation by a single parity predicate, and encoding this
predicate by a circuit gives exactly \eqref{eq:source}. Conversely, a
$\Sig$ machine can make the existential and universal choices and then
ask a single parity query. Hence the quantified circuit problem is
complete for $\Sig$.

Membership of BCD in $\Sig$ follows from the same form. With $n$ declared
inputs, we guess an $n$-bit indicator $s$ of a set $S$ with
$1\le|S|\le k$. We then universally choose $y\in\{0,1\}^n$, which
supplies the inputs outside $S$. Finally, we take the parity over
$z\in\{0,1\}^n$ of
\[
 \mathbf1[\,z_i=0\text{ whenever }s_i=0\,]\;f(x),
 \qquad x_i=\begin{cases}z_i&s_i=1,\\y_i&s_i=0.\end{cases}
\]
Notice that only the coordinates selected by $s$ are allowed to vary in
$z$. Every assignment to $S$ therefore contributes exactly once, and the
parity is $\D_Sf(y)$. Requiring this value to be one for every $y$ is
exactly the certificate condition.

\section{Further complexity results for zero cubes}\label{app:zero}
In \cref{sec:zero}, we proved the parameterized gap for zero cubes. In
this appendix, we derive its approximation consequences, and we then
classify zero cubes in the classical and fixed threshold settings. Our
last construction keeps the formula satisfiable while making its
satisfying assignments far apart, which yields stronger approximation
bounds when clauses of size three are allowed.

\subsection{The approximation bounds}\label{app:zero-approx}
\begin{proof}[Proof of \cref{cor:zero-approx}]
Suppose that an algorithm achieves the approximation factor
$\rho(\zeta(F))$ in time $h(\zeta(F))\ell^{O(1)}$. We choose
\[
 g(k)=\max\{k,\max_{1\le j\le k}j\rho(j)\},
\]
and use the reduction of \cref{thm:zero-gap} with its odd $L>g(k)$. If
the optimum is some $j\le k$, the algorithm must return a set of at most
$j\rho(j)\le g(k)<L$ variables. If the optimum is $L$, every feasible
answer has at least $L$ variables. Hence the size of the returned set
distinguishes the two cases.

We must also bound the running time in both cases. The supplied zero
cube of size $L$ guarantees $1\le\zeta(F)\le L$, so the time is at most
\[
 \left(\max_{1\le j\le L}h(j)\right)\ell^{O(1)}.
\]
Since $L$ depends only on $k$, this would be an FPT algorithm for the
complement of Multicolored Clique, and it would imply FPT=W[1]. Note that
taking the maxima above avoids any assumption that $h$ or $\rho$ is
monotone.

For ordinary polynomial time approximation, we start from a
Multicolored Clique instance with $v$ vertices in total, and choose an
odd $L=\Theta(v^B)$ for a sufficiently large fixed constant $B$. The
small optimum is at most $r\le v$, whereas the large optimum is $L$. The
output has
\[
 N=O(vL),\qquad \ell=O(v^2L^2\log(vL)).
\]
For the two approximation factors, the ratio between the largest answer
permitted in the small case and the optimum in the large case is bounded
by
\[
 \frac{rN^{1-\varepsilon}}L
   =O(v^{2-\varepsilon}L^{-\varepsilon}),\qquad
 \frac{r\ell^{1/2-\varepsilon}}L
   =O\bigl(v^{2-2\varepsilon}L^{-2\varepsilon}
                (\log(vL))^{1/2-\varepsilon}\bigr).
\]
For each permitted $\varepsilon$, both expressions tend to zero once $B$
is large enough. Such an approximation would therefore distinguish the
two cases, and decide Clique in polynomial time. Since $B$ is fixed, the
construction remains polynomial, and the finitely many small source
instances can be decided directly.
\end{proof}

\subsection{Classical complexity and fixed thresholds}
We now turn to the classical setting, in which the cube size is part of
the input. The theorem below collects all of our results for zero cubes
in this setting.

\begin{theorem}\label{thm:zero-classical}
Deciding $\zeta(F)\le k$ is $\Sig$-complete on positive unit-free $2$CNF.
For every fixed $k\ge1$, it is polynomial-time decidable on signed $2$CNF,
and coNP-complete on $3$CNF. On satisfiable, unit-free, all-essential
$3$CNF, coNP-completeness holds for every fixed $k\ge2$. On the same class,
for every fixed $0<\varepsilon<1$, no polynomial-time $N^{1-\varepsilon}$
or $\ell^{1-\varepsilon}$ approximation exists unless P=NP.
\end{theorem}

\begin{proof}[Proof of classical completeness]
Membership follows directly from the definition. We guess a nonempty set
of at most $k$ coordinates, universally choose an assignment to the other
coordinates, and use a single parity query to test whether the number of
satisfying completions is even.

For hardness, we start from the quantified parity
problem~\eqref{eq:source}. We would like the choice of a cube to amount
to the choice of an existential assignment $\alpha$. To do so, we
introduce a pair of variables $a_i,b_i$ for each existential bit $u_i$,
where choosing $a_i$ represents $\alpha_i=1$ and choosing $b_i$
represents $\alpha_i=0$. Let $Z$ be the set of all coordinates of $u$ and
$w$, and form the circuit
\[
 C=(R(u,v,w)+\mathbf1[w=0])
       \prod_i\bigl(u_ia_i+(1+u_i)b_i\bigr).
\]
For a fixed $u=\alpha$, the product is the monomial $x_{A_\alpha}$ that
uses exactly the chosen member of every pair. Summing over $u$ and $w$
therefore gives
\[
 \D_ZC=\sum_\alpha x_{A_\alpha}(h_\alpha(v)+1).
\]
The added indicator contributes one to the parity sum, also when $w$ is
empty. In this way the condition $h_\alpha\equiv1$ becomes exactly the
zero derivative we want.

We still have to force a small cube to include $Z$ and one member of each
pair. To do so, we add an unused coordinate $e$, let $V'$ be the set of
all coordinates including $e$, and put
\[
 G=C+\sum_{z\in Z}x_{V'\setminus\{z\}}
       +\sum_i x_{V'\setminus\{a_i,b_i\}}.
\]
Since every added term contains $e$, none of them can cancel with a term
of $C$, and the added terms are also distinct from each other. A zero
cube must avoid being contained in any of them. The term that misses only
$z$ therefore forces the cube to contain $z$, and the term that misses
only $a_i,b_i$ forces it to contain a member of that pair. At budget
$K=|Z|+p$, these requirements leave exactly the sets $Z\cup A_\alpha$.
Their derivatives kill every added term, together with the monomial of
every other assignment, which leaves $h_\alpha+1$. Hence $\zeta(G)\le K$
exactly when the quantified parity instance is positive. Notice also that
$G\ne0$, and that its full monomial is absent.

To obtain a positive $2$CNF, we apply the compiler and the crown
construction to $G$, treating every coordinate of $V'$, those of $Z$
included, as a public input of \cref{lem:compiler}, so that no circuit
inputs are summed out. Each guard of \cref{lem:crown} misses just one
coordinate of $M\cup T$, so a zero cube must contain all of $M\cup T$.
Once it does, the guards disappear from the sum. What remains is zero
exactly when the public part of the cube is a zero cube of $G$, or when
the cube contains the unused coordinate $d$ of the crown. Using
\cref{lem:zero-block}, we give $d$ an odd weight $L>K$, and we give every
other coordinate weight one. The new optimum is
\[
 2m+\min\{\zeta(G),L\}.
\]
The threshold $2m+K$ therefore proves hardness on positive $2$CNF. Note
that selecting $M\cup T$ together with the whole block of $d$ always
supplies a feasible zero cube.
\end{proof}

\begin{proof}[Proof of the fixed-threshold claims without promises]
For signed $2$CNF and a fixed $k$, we enumerate all nonempty sets $S$ of
at most $k$ variables. Fixing $S$ in all $2^{|S|}$ ways gives a list of
cofactors, each of which is a $2$CNF on the remaining variables. The
derivative is nonzero exactly when some outside assignment makes an odd
number of these cofactors true.

We can check this by enumerating the patterns of true and false values
with an odd number of trues. For a given pattern, we conjoin all
cofactors that are required to be true. For each cofactor required to be
false, we choose one of its clauses to falsify, and add the corresponding
unit requirements, of which there are at most two. The resulting formula
is still a $2$CNF, so $2$SAT decides whether the pattern can occur.
Constant cofactors are handled directly. For a formula with $m$ clauses
there are at most $2^{2^k}(m+1)^{2^k}$ choices, which is polynomial for
fixed $k$.

On circuits, the problem lies in coNP for each fixed $k$. To certify that
no candidate $S$ is a zero cube, it suffices to give, for every
candidate, an outside assignment at which its derivative is odd. There
are only polynomially many candidates, and checking each of them uses at
most $2^k$ circuit evaluations.

For hardness on $3$CNF, we take a $3$CNF source and replace each of its
variables by $k+1$ copies, which binary clauses force to be equal. Each
original occurrence is kept on one representative copy. If the source is
unsatisfiable, the resulting function is zero, and every set is a zero
cube. If the source is satisfiable, any two distinct satisfying
assignments differ in at least $k+1$ coordinates. Given any proposed cube
$S$ of at most $k$ coordinates, we fix the outside coordinates as in a
satisfying assignment. The cube then contains exactly one satisfying
assignment, so its derivative is one at that point, and $S$ is not a
zero cube. This is a reduction from unsatisfiability, for which a source
with at least one variable suffices.
\end{proof}

\subsection{Keeping the formula satisfiable and every variable essential}
The reduction above uses an unsatisfiable formula for its small optimum
case, and we now remove that feature. Our new construction always has two
satisfying assignments that are close to each other. Whereas these two
exist regardless of the source, a satisfiable source circuit also
produces satisfying assignments that are far from every other satisfying
assignment. These isolated assignments are what prevent small zero
cubes.

\begin{proof}[Proof of the promised bounds in \cref{thm:zero-classical}]
Let $C$ be a circuit of size $s$, and let $L\ge3$ be given in unary. We
first simplify constant gates, as well as AND and OR gates with repeated
inputs. If this leaves a constant circuit, we replace it by $x$ when it
is satisfiable and by $x\wedge\neg x$ otherwise. We then encode the gate
truth tables by a unit-free $3$CNF $K(y,w)$, with inputs $y$ and wire
variables $w$, and leave the output $o$ unpinned. Note that each input
has a unique wire assignment.

We add a switch $t$ and the clauses
\[
 (\neg t\vee o),\qquad (t\vee\neg y_i)\quad\text{for every input }y_i.
\]
When $t=0$, all inputs are zero, which gives one state of the inputs and
wires. When $t=1$, the output must be one, so the states correspond to
the satisfying inputs of $C$. We then replace every input, wire and
switch variable by $L$ copies, using binary clauses to force them to be
equal, and keep the old occurrences on a representative. As a result,
distinct states differ in at least $L$ coordinates.

Next, we give each state two control assignments. We introduce
$a_1,\ldots,a_L$, impose $a_1=a_2$, and for $j\ge3$ impose
\[
 (t\vee\neg a_j),\quad
 (\neg t\vee\neg a_1\vee a_j),\quad
 (\neg t\vee a_1\vee\neg a_j).
\]
At $t=0$, the controls are $0^L$ or $110^{L-2}$, which gives two
satisfying assignments at distance two, regardless of $C$. At $t=1$, the
controls are $0^L$ or $1^L$. Hence every satisfying assignment with $t=1$
is at distance at least $L$ from every other satisfying assignment:
changing the state changes an entire block of copies, while changing only
the controls changes all $L$ of them.

Notice that the whole control block is always a zero cube, since every
outside assignment has zero or two satisfying extensions. If $C$ is
unsatisfiable, only the two assignments at $t=0$ remain. In this case
$\{a_1,a_2\}$ is a zero cube, whereas no singleton is one, because a cube
of one coordinate through either satisfying assignment contains only that
assignment. Thus $\zeta=2$.

If $C$ is satisfiable, we take a satisfying assignment with $t=1$. Every
cube of fewer than $L$ coordinates through it contains exactly one
satisfying assignment, so no such set can be a zero cube, and $\zeta=L$.
Finally, in both cases every satisfying assignment has distance at least
two from every other one. Flipping a single coordinate of a satisfying
assignment therefore makes the formula false, and consequently every
coordinate is essential.

The construction uses $O(sL)$ variables and clauses, so
$\ell=O(sL\log(sL))$. Taking $L=k+1$ proves the promised coNP-hardness
for each fixed $k\ge2$, and membership was proved above. For
approximation, we take $L=s^B$ with a fixed integer $B>1/\varepsilon$.
Then
\[
 \frac{2\ell^{1-\varepsilon}}L
 =O\bigl(s^{1-\varepsilon-B\varepsilon}
              (\log s)^{1-\varepsilon}\bigr)\longrightarrow0.
\]
An approximation with this factor must return fewer than $L$
coordinates in the unsatisfiable case, whose optimum is two, whereas every
feasible answer in the satisfiable case has at least $L$ coordinates.
Such an approximation would therefore decide satisfiability in polynomial
time. The same calculation with $N=O(sL)$ proves the $N^{1-\varepsilon}$
bound, and this completes the proof of \cref{thm:zero-classical}.
\end{proof}

\section{Reducing signed 2CNF to a small equivalent instance}\label{app:kernel}
In this appendix, we prove the kernel claimed in part (a) of
\cref{thm:main-separation}. The algorithm of \cref{sec:small-one}
searches among representatives, but still checks every candidate against
the original formula. A kernel, on the other hand, has to justify
deleting the other variables altogether. Although we would normally
expect $k$ copies of each type to be enough, they are not -- a candidate
could select all of them, which would remove the assignments that the
deleted copies used to supply. We therefore keep two extra copies of
each type, so that this never happens.

\begin{theorem}\label{thm:kernel}
BCD on signed $2$CNF has a polynomial-time kernel with at most
\[
 2k+(k+2)16^{2k}
\]
variables and $2^{O(k)}$ encoding length.
\end{theorem}
\begin{proof}
We use the preprocessing, the cover $T$ and the types from the proof of
\cref{thm:small-one}, and replace an instance rejected there by a fixed
NO instance. For every remaining type, we keep at most $k+2$ variables,
and delete the others together with their incident clauses. We now show
that the resulting formula has a certificate of each size at most $k$
exactly when the original formula does.

To do so, we first leave all cover variables unassigned, and write $t$
for them. For one outside type, let $q(t,x)$ be the conjunction of the
clauses joining a representative $x$ to the cover, and write
$q_b(t)=q(t,b)$. Each copy of the type contributes the same factor $q$ on
its own variable. Suppose that a proposed certificate selects $r$ copies.
Summing over them contributes the factor
\[
 (q_0+q_1)^r=
 \begin{cases}1&r=0,\\q_0+q_1&r>0.\end{cases}
\]
The equality holds because a Boolean valued function satisfies $a^r=a$
for every positive integer $r$.

Now consider the $u$ copies that are left outside the proposed
certificate. Their assignments produce the following possible functions
of the cover:
\[
 \begin{array}{c|c}
 u&\text{possible products}\\\hline
 0&\{1\}\\
 1&\{q_0,q_1\}\\
 u\ge2&\{q_0,q_1,q_0q_1\}.
 \end{array}
\]
This is because with at least two copies, an assignment is all zero, all
one, or contains both values, and repeating a factor has no effect. Two
unselected copies thus already supply every product that a larger number
of copies can supply. It is salient to point out that we compare these
products as functions of $t$, rather than value by value. This matters
because some cover variables may later be summed over, and the outside
assignment has to remain fixed during that sum.

If a type originally had more than $k+2$ copies, a selection of at most
$k$ variables necessarily leaves at least two of them unselected, both
before and after the deletion. We can therefore keep the same selected
count $r$. Its summed factor is unchanged, and by the table, its
unselected copies provide exactly the same family of functions of $t$.

Since the choices for different types are independent, multiplying these
factors with the clauses contained entirely in $T$ gives the same family
of functions on $T$ for both formulas. Finally, we sum over the selected
cover variables. In either formula, every remaining assignment gives one
exactly when the same holds in the other, and this is the certificate
condition. Matching the selected counts within each type preserves the
size of the certificate in both directions.

Partitioning the variables into types and deleting the excess copies
takes polynomial time. There are at most $2k$ cover variables and at most
$16^{2k}$ types, which gives the stated bound on variables. Further, once
repeated clauses are removed, the number of signed binary clauses is at
most quadratic in the number of remaining variables. Relabeling the
variables therefore gives an encoding of $2^{O(k)}$ bits.
\end{proof}

\section{Proofs for structural width}\label{app:width}
In this appendix, we prove the treewidth results summarized in
\cref{sec:other}. We begin by describing the information that the dynamic
program stores at a bag, and we show why it suffices to store its affine
hull. We then give the dynamic program itself. Lastly, we show that for
circuits the doubly exponential dependence on the width is necessary,
unless $\mathrm{NP}\subseteq\mathrm{SUBEXP}$.

\subsection{The information stored at a separator}
We write the input as a product of local Boolean factors,
\[
 F(x)=\prod_j\phi_j(x|_{U_j}).
\]
Our algorithm allows some variables to be summed in advance. We call this
set $A$ mandatory, and we call a second set $E$, disjoint from $A$,
eligible. Selection of all remaining variables is forbidden. The task is
to minimize $|S|$ over nonempty $S\subseteq E$ for which
$\D_{A\cup S}F\equiv1$. Note that ordinary BCD is the case
$A=\varnothing$ and $E=V$.

\begin{theorem}\label{thm:algorithm}
Given a width-$w$ tree decomposition of the factor primal graph, this optimum
and a minimizing set can be found in $2^{2^{O(w)}}\poly(\ell)$ time, or
infeasibility can be reported. Here $\ell$ includes the factors and the
decomposition.
\end{theorem}

We root the decomposition and consider a bag $B$. The variables that
occur below $B$ but not in $B$ are its internal variables, and we fix a
selection among them. For each assignment $y$ to the internal unselected
variables, we sum the product of the factors below $B$ over the selected
and mandatory internal variables. What remains is a function of the bag
variables, or equivalently a vector
\[
 r_y\in\F^{2^{|B|}}
\]
whose entries are indexed by the assignments to $B$. We call it a
response table. The family $R=\{r_y:y\}$ is nonempty, even when every
table in it is zero. Observe that the selection stays fixed while $y$
varies.

The dynamic program needs whole tables, rather than information kept
entry by entry. For example, consider: $B$ consists of a single variable,
and the two families are
\[
 \{(1,0),(0,1)\}\quad\text{and}\quad
 \{(0,0),(0,1),(1,0),(1,1)\}.
\]
Both families allow both values in each entry, so information kept entry
by entry cannot distinguish them. However, summing out the bag variable
gives one on every table of the first family, whereas the second family
also produces zero.

Keeping the exact family would be costly. Instead, we store its affine
hull, which over $\F$ consists of all linear combinations of members of
the family whose coefficients sum to one. The next lemma shows that this
compression is safe.

\begin{lemma}\label{lem:affine}
For nonempty vector families, linear maps $L,L_c$, and a bilinear map $\beta$,
\begin{align*}
 \aff(L(R))&=L(\aff(R)),\\
 \aff\left(\bigcup_cL_c(R)\right)
 &=\aff\left(\bigcup_cL_c(\aff(R))\right),\\
 \aff(\beta(R_1,R_2))
 &=\aff(\beta(\aff(R_1),\aff(R_2))).
\end{align*}
\end{lemma}
\begin{proof}
A linear map takes an affine combination to the same combination of the
images, which proves the first identity. Applying it to each $L_c$, and
then taking the affine hull of the union, proves the second.

For the third, let $\sum_i\lambda_i=\sum_j\mu_j=1$. Bilinearity gives
\[
 \beta\left(\sum_i\lambda_i r_i,\sum_j\mu_j s_j\right)
 =\sum_{i,j}\lambda_i\mu_j\beta(r_i,s_j),
 \qquad \sum_{i,j}\lambda_i\mu_j=1.
\]
Combining two points from the affine hulls thus produces an affine
combination of the original products, which gives one containment. The
other follows because each original family is contained in its affine
hull.
\end{proof}

\subsection{The dynamic program}
\begin{proof}[Proof of \cref{thm:algorithm}]
We use a rooted nice tree decomposition whose root and leaves are empty,
and assign each factor to one bag that contains all of its variables.
Such a bag exists, since these variables form a clique in the primal
graph.

For each bag, a state consists of an affine space of response tables
together with a bit, which records whether an eligible variable has been
selected. For each state, we store the smallest cost of a selection that
produces it. The bit is needed because the final selection must be
nonempty, and summing mandatory variables alone does not meet that
requirement. At an empty leaf, the response space is $\{1\}$, the bit is
false, and the cost is zero.

It is easiest to describe the transitions on exact families first, and
then to apply \cref{lem:affine} so as to use affine hulls instead.
\begin{itemize}
\item When a variable is introduced, we extend each table so that it does
not depend on the new coordinate. When a factor is introduced, we
multiply each table pointwise by the factor.
\item At a join, we take every pointwise product of a table from the
first child with a table from the second. We add their costs, and take
the OR of their bits.
\item When we forget a selected variable $v$, we replace each table $r$
by $r(\cdot,0)+r(\cdot,1)$, add one to the cost, and set the bit to true.
\item When we forget an unselected variable, we keep both restrictions
$r(\cdot,0)$ and $r(\cdot,1)$ of each table $r$, which forms the union of
the two restricted families.
\item A mandatory variable uses the summed transition at zero cost and
leaves the bit unchanged. A forbidden variable uses the unselected
transition, and an eligible variable is allowed both choices.
\end{itemize}

We observe that these transitions preserve the meaning of a response
family. At a join, the internal variables of the two children are
disjoint. Their unselected assignments therefore vary independently, and
the parity sums over their selected variables multiply. Further, a bag
variable is selected only at its unique highest forget node, so its
occurrences in different children cannot lead to conflicting choices.
When a forgotten variable is summed, its two values are added within the
same table, in which all other unselected variables have the same
assignment.

Every transition is a linear map, a union of linear images, or a
bilinear product. By \cref{lem:affine}, applying it to affine hulls gives
the correct affine hull of the new response family. It follows by
induction that two selections with the same state and bit are
interchangeable in all later computations, and we can discard the more
expensive one.

At the empty root, no bag variables are left, so each response is a
single bit. A nonempty family of bits has affine hull $\{1\}$ exactly
when every response is one, which is the certificate condition. We
therefore accept the affine space $\{1\}$ with the bit set to true.

We now bound the number of states and the cost of updating them. We put
$d=2^{|B|}$. Since an affine space in $\F^d$ can be specified by a base
vector and at most $d$ generators, there are at most $2^{d^2+d}$
possibilities. Gaussian elimination gives a canonical representation, in
which the linear transitions are immediate. For the join, suppose that
the spaces are $v+U$ and $z+W$, with bases $(u_i)$ and $(w_j)$. Their
product hull is
\begin{equation}\label{eq:affinejoin}
 v\circ z+\spanf\{u_i\circ z,\ v\circ w_j,\ u_i\circ w_j:i,j\},
\end{equation}
where $\circ$ denotes pointwise multiplication. Expanding
$(v+u)\circ(z+w)$ proves containment in the displayed affine space.
Conversely, $u_i\circ z$ and $v\circ w_j$ are differences of two
products, and $u_i\circ w_j$ is a sum of four products, so every
displayed generator lies in the direction space of the product hull.

Each update takes time polynomial in $d$ for a state, or for a pair of
states. Since $|B|\le w+1$, the total running time is
\[
 2^{O(d^2)}\poly(\ell)=2^{2^{O(w)}}\poly(\ell).
\]
Storing the choices that produced each retained entry recovers a minimum
set.
\end{proof}

For zero cubes, we change only the accepting state at the root, to
$\{0\}$. The same argument then shows that every outside assignment has
even parity. Here too the algorithm allows mandatory, eligible and
forbidden variables.

\begin{corollary}\label{cor:circuits}
For a fan-in-two circuit, BCD is solvable in
$2^{2^{O(w)}}\poly(\ell)$ time given a width-$w$ decomposition of its
functional gate-constraint primal graph.
\end{corollary}
\begin{proof}
We give every gate a wire variable and a factor enforcing its truth
table, and add a factor requiring the output to be one. For a fixed
input, exactly one wire assignment satisfies all gate factors, so summing
the product over the wires gives the value of the original circuit. We
declare the wires mandatory at zero cost. In this setting, only the
selected input variables contribute to the cost and to the bit for a
nonempty selection, and we can apply \cref{thm:algorithm}.
\end{proof}

\subsection{The lower bound for circuits}
We now show that for circuits the doubly exponential dependence on the
width cannot be improved, unless $\mathrm{NP}\subseteq\mathrm{SUBEXP}$.
The difficulty is that a bag of small width cannot hold an entire SAT
instance. Our reduction therefore uses an address, and the circuit
examines only the clause named by that address. Since an assignment must
satisfy every clause, we need a universal quantification over addresses,
and the requirement of a constant derivative supplies it.

\begin{theorem}\label{thm:eth}
Unless $\mathrm{NP}\subseteq\mathrm{SUBEXP}$, circuit BCD has no
$2^{2^{o(w)}}\poly(\ell)$ algorithm when $w$ is the pathwidth of its
functional gate-constraint primal graph, even with a decomposition supplied.
The same lower bound therefore holds for treewidth.
\end{theorem}
\begin{proof}
We start with $3$SAT on $n$ variables. We remove duplicate and
tautological clauses, and handle an empty clause, or fewer than two
remaining clauses, directly. There are then $m=O(n^3)$ clauses, and we
duplicate clauses until their number is a power of two. An address $y$
of $q=O(\log n)$ bits then selects one clause. For a variable $i$ and a
value $b$, we define $g_i^b(y)$ to be zero if assigning the value $b$ to
variable $i$ satisfies a literal of the addressed clause, and one
otherwise.

We represent the two possible values of source variable $i$ by a pair of
new variables $a_i,b_i$, and consider the Boolean function
\begin{equation}\label{eq:ethcircuit}
 F(a,b,y)=\prod_i(a_i+b_i)
       +\prod_i\bigl(a_i g_i^1(y)+b_i g_i^0(y)\bigr).
\end{equation}
For a source assignment $\alpha$, let $A_\alpha$ contain $a_i$ when
$\alpha_i=1$, and $b_i$ when $\alpha_i=0$. Expanding the products gives
\[
 F=\sum_\alpha x_{A_\alpha}h_\alpha(y),\qquad
 h_\alpha(y)=1+\prod_i g_i^{\alpha_i}(y).
\]
The product is one exactly when no variable satisfies the addressed
clause under $\alpha$. Hence $h_\alpha(y)$ is the indicator that $\alpha$
satisfies the addressed clause.

Every $A_\alpha$ has size $n$, and different choices are incomparable. A
monomial arising from one assignment therefore cannot contain a monomial
arising from another. For a fixed $\alpha$, the maximal supports are the
sets $A_\alpha\cup J$ with $J\in\Max(h_\alpha)$. Such a support has size
at most $n$ exactly when $J=\varnothing$ is maximal, i.e., when
$h_\alpha\equiv1$. By \cref{lem:anf},
\[
 \depth(F)\le n
 \quad\Longleftrightarrow\quad
 \text{the source formula is satisfiable}.
\]

Next, we implement \eqref{eq:ethcircuit} with small pathwidth. The
function $g_i^b$ is one plus the XOR of the address minterms of the
clauses that contain a literal on variable $i$ satisfied by the value
$b$. Since these minterms are disjoint, their XOR is the indicator of
those clauses. We compute each minterm with an AND chain and add it to
an XOR accumulator, using new gates for different uses. The two products
in \eqref{eq:ethcircuit} are computed sequentially over $i$.

Our path decomposition keeps the $q$ address inputs throughout. It also
keeps the two product accumulators, the current pair $a_i,b_i$, and a
constant number of temporary wires. For each gate, we introduce its
output before forgetting any of its inputs at their last use. The scope
of each gate then lies in a bag, and each variable occurs in an interval
of bags. Hence the width is $q+O(1)$, and the circuit has polynomial
size. Note that replacing the XOR gates by constant size circuits of AND,
OR and NOT gates changes neither bound asymptotically.

A hypothetical $2^{2^{o(w)}}\poly(\ell)$ algorithm, applied with this
decomposition, would therefore solve $3$SAT in time
\[
 2^{2^{o(\log n)}}n^{O(1)}=2^{n^{o(1)}}.
\]
In that case $3$SAT, and with it every problem in NP, would be in
SUBEXP.
\end{proof}

\section{The clause-width boundary for constant one}\label{app:horn}
In this appendix, we ask how much the $2$CNF assumption matters for
finding a small certificate. We show that increasing the clause width
from two to three changes the parameterized problem dramatically.
Strikingly, this happens even for Horn formulas, whose satisfiability
problem is decidable in polynomial time.

\begin{theorem}\label{thm:horn}
For every computable $g:\mathbb N\to\mathbb N$ with $g(k)\ge k$,
distinguishing $\depth(F)\le k$ from $\depth(F)>g(k)$ is coW[1]-hard
parameterized by the supplied threshold $k$. This holds on satisfiable,
unit-free Horn $3$CNFs with every coordinate essential and only one variable
occurring positively.
\end{theorem}
\begin{proof}
We reduce from the complement of Multicolored Clique, and our
construction has three parts. Firstly, selectors make a derivative
recognize a multicolored clique. Secondly, a switch changes this
derivative to its complement. Thirdly, blocks make every alternative
certificate too large.

Take a graph with nonempty color classes $V_1,\ldots,V_r$, where $r\ge2$.
We call two distinct vertices incompatible if they belong to the same
class, or form a nonedge between different classes. A variable $y_v$
records whether vertex $v$ is selected, so that
\[
 B(y)=\bigwedge_{\{u,v\}\text{ incompatible}}(\neg y_u\vee\neg y_v)
\]
allows at most one vertex per class, and requires the selected vertices
to be pairwise adjacent. However, it does not yet require a vertex from
every class. For that purpose, we introduce $a_1,\ldots,a_r$ and set
\[
 Q(a,y)=B(y)\wedge\bigwedge_{i,\,v\in V_i}(a_i\vee\neg y_v),
 \qquad A=\{a_1,\ldots,a_r\}.
\]
Fix $y$. If class $i$ contains a selected vertex, its clauses force
$a_i=1$, and summing over $a_i$ gives one. If no vertex of the class is
selected, either value of $a_i$ is allowed, and the sum gives zero.
Consequently,
\begin{equation}\label{eq:horn-clique}
 \D_AQ=B(y)\prod_i\left(\bigvee_{v\in V_i}y_v\right)=:C_G(y).
\end{equation}
The function $C_G$ is the indicator of multicolored cliques. In
particular, it is identically zero exactly when there is no such clique.

In that case, we need a derivative equal to one. To do so, we add a
switch $s$ and let
\[
 H=\bigwedge_{C\in Q}(s\vee C)
       \wedge\bigwedge_i(\neg s\vee a_i).
\]
At $s=0$ the formula is $Q$, and at $s=1$ it is $\prod_i a_i$. Taking the
parity of these two cases gives
\begin{equation}\label{eq:hornswitch}
 H=Q+sQ+s\prod_i a_i,
 \qquad \D_{\{s\}\cup A}H=1+C_G(y).
\end{equation}
Hence $\{s\}\cup A$ is a certificate exactly when there is no clique.
Note also that $Q(a,0)=1$. Setting all vertex variables to zero in the
displayed formula for $H$ therefore gives $1+s+s\prod_i a_i$, and the
only ANF supports of $H$ that avoid the vertex variables are
$\varnothing$, $\{s\}$ and $\{s\}\cup A$.

We still have to exclude other small certificates. We put $k=r+1$ and
$L=g(k)+1$, and replace every $y_v$ by the OR of $L$ new bits. We claim
that a maximal monomial that uses any bit of a block must use the whole
block. To see this, write
\[
 \bigvee_{j=1}^Lz_j=\sum_{\varnothing\ne J\subseteq[L]}z_J.
\]
In the substituted polynomial, the blocks used by a term identify the
original vertex variables in that term, and together with its unchanged
variables they identify the original support. Different original terms
therefore cannot cancel. Further, every nonempty subset of a used block
occurs, so a term that uses only part of a block lies inside a larger
term that uses all of it, and cannot be maximal.

Consequently, every certificate of size less than $L$ must avoid the
blocks, and of the three supports listed above, only $\{s\}\cup A$ can be
maximal. The OR substitution preserves whether $C_G$ is zero, because
every assignment to the old vertex variables is realized by an
assignment to the blocks. By \eqref{eq:hornswitch}, the optimum is
therefore $k$ when there is no clique, and at least $L>g(k)$ when there
is one.

We finish by checking the promised class of formulas. Every occurrence
of a vertex variable in $H$ is negative, so replacing it by an OR and
distributing the clauses keeps the width at most three, with
$O(L^2|V(G)|^2)$ clauses. We complement all selectors $a_i$. Before the
block substitution, the clauses then take the forms
\[
 (s\vee\neg y_u\vee\neg y_v),\qquad
 (s\vee\neg a_i\vee\neg y_v),\qquad
 (\neg s\vee\neg a_i).
\]
After substitution and distribution they remain Horn and unit-free, with
only $s$ occurring positively. Recall that complementing inputs
preserves certificates.

Satisfiability and essentiality are easiest to check before this last
complementation. Taking $s=1$ and all $a_i=1$ satisfies the formula. At
$s=1$, changing any selector from one to zero changes the output. At
$y=0$ with one selector at zero, changing $s$ changes the output. For a
bit in the block of a vertex in class $i$, we take $s=0$, $a_i=0$, all
other selectors one, and all other block bits zero. Changing that bit
from zero to one then changes the output. Every variable is therefore
essential. Finally, $L$ depends only on $r$, so the construction is an
FPT reduction.
\end{proof}

\subsection{Fixed thresholds}\label{app:fixed}
\Cref{thm:horn} rules out an FPT algorithm under the usual assumption.
However, it does not rule out a polynomial time algorithm for each fixed
$k$. Whereas at width three such an algorithm exists, at width four even
this weaker possibility fails unless P=NP.

\begin{proposition}\label{prop:fixed}
For every fixed $k\ge1$, BCD is polynomial-time decidable on $3$CNF and
coNP-complete on $4$CNF.
\end{proposition}
\begin{proof}
For $3$CNF, we enumerate the nonempty sets $S$ of at most $k$ variables,
after discarding tautological clauses. If $S$ misses a remaining clause,
it is not a certificate. Otherwise, fixing the variables of $S$ leaves a
$2$CNF, since every surviving clause has lost at least one literal. There
are at most $2^k$ such cofactors, one for each assignment to $S$.

The derivative is the parity of these cofactors. To disprove that it is
always one, we must find an outside assignment that makes an even number
of cofactors true. We therefore enumerate all even patterns of true and
false values. For a chosen pattern, we include every clause of each
cofactor that is required to be true. For each cofactor required to be
false, we choose one of its clauses to falsify, which amounts to
imposing at most two unit requirements. The resulting test is a $2$SAT
instance.

Constant cofactors are handled directly, since a constant cofactor is
consistent only with its own value and requires no clause choice. There
are at most $2^{2^k}$ patterns and $(m+1)^{2^k}$ clause choices per
pattern, where $m$ is the number of input clauses, which is polynomial
for fixed $k$. An attainable even pattern rejects $S$, and if none
exists, $S$ is a certificate.

For width four, we first take $k=1$. Given a $3$CNF $F(y)$, we introduce
new variables $x,u,v$ and form
\[
 G(x,u,v,y)=\bigwedge_{C\in F}(x\vee C)\wedge(x\vee u)\wedge(x\vee v).
\]
Observe that the last two clauses force any singleton certificate to be
$\{x\}$. At $x=1$ the formula is one, and at $x=0$ it is $F(y)uv$, so
$\D_{\{x\}}G=1+F(y)uv$, which is identically one exactly when $F$ is
unsatisfiable. For a larger fixed $k$, we conjoin $k-1$ new unit
variables. Every maximal ANF support then gains exactly those variables,
so a certificate of size at most $k$ exists exactly when $G$ has a
singleton certificate. This proves coNP-hardness.

For membership in coNP, a NO instance can give, for each candidate set,
one outside assignment with derivative zero. There are polynomially many
candidates when $k$ is fixed, and each derivative value can be checked
with at most $2^k$ evaluations of the input formula.
\end{proof}

\subsection{Checking a specified set}
The same cofactor argument also classifies the task of checking one
proposed certificate, where the set is supplied and its size is the
parameter.

\begin{proposition}\label{prop:verifier}
Given a $3$CNF $F$ and a specified nonempty $S$, deciding whether
$\D_SF\equiv1$ is coW[1]-complete parameterized by $|S|$. Hardness already
holds for the unamplified Horn construction in \cref{thm:horn}.
\end{proposition}
\begin{proof}
For hardness, we use \eqref{eq:hornswitch}, complement the selectors as
above, and prescribe $S=\{s,a_1,\ldots,a_r\}$. The specified set is a
certificate exactly when the source graph has no multicolored clique.
Note that no block amplification is needed here.

For membership, we reduce the complementary problem to Clique. We first
remove tautological clauses. If $S$ misses a remaining clause, then $S$
is immediately not a certificate, and we output a fixed YES instance of
Clique. Otherwise, we form the cofactors of width two and enumerate the
even patterns as in the preceding proof. Fix one pattern. Its true
cofactors give a base $2$CNF $B$, and if $B$ is unsatisfiable we discard
the pattern. Every nonconstant cofactor required to be false asks us to
choose a clause to falsify, and each such choice is a bundle of at most
two unit assumptions. Constant cofactors are handled as before.

These choices can be encoded by a graph, because of the following
property. A satisfiable $2$CNF together with unit assumptions is
satisfiable whenever each assumption, and each pair of assumptions, is
consistent with the base formula. To see this, we existentially
eliminate, by resolution, all variables not fixed by the assumptions.
The resulting formula on the fixed variables is still a $2$CNF. If the
assumed values were inconsistent with it, they would falsify a clause of
size at most two, or the assumptions themselves would contain a
contradictory pair. Either failure would already be visible in one or two
assumptions.

We make one vertex group for each cofactor that requires a clause
choice, and one vertex for each clause it could falsify. We delete the
choices whose bundle is inconsistent with $B$, and we join vertices from
different groups exactly when their two bundles are jointly consistent
with $B$. A clique that chooses one vertex from every group is then a set
of pairwise consistent choices. By the property above, these choices are
jointly consistent, and hence give an assignment that realizes the
pattern. Conversely, such an assignment falsifies a clause of every false
cofactor, and these clauses form a clique.

There are at most $t=2^{|S|}$ groups. We add singleton groups, whose
vertex is adjacent to every vertex of every other group, until there are
exactly $t$ groups. A pattern that requires no choices then produces a
$t$-clique, as needed, since its base is satisfiable. We take the
disjoint union of these graphs over all patterns that were not
discarded. Since $S\ne\varnothing$, we have $t\ge2$, and a $t$-clique must
lie in one component. The final graph therefore has a $t$-clique exactly
when some even pattern is attainable. Constructing it takes FPT time, and
$t$ depends only on $|S|$.
\end{proof}

\section{Implications and the binary-language dichotomy}\label{app:implication}
In this appendix, we prove the dichotomy for binary constraint languages
stated in \cref{sec:other}. The only hard case that needs new work is
implication. An implication $x\to y$ forbids just the assignment
$x=1,y=0$, and we use this simple constraint to obtain the same family of
certificates as in the lifting theorem for positive $2$CNF. The
difficulty is that an implication treats its two ends differently. To
satisfy many implications at once, we can either set their tails to zero
or set their heads to one, and our construction uses both of these moves.

\begin{theorem}[Implication lifting]\label{thm:implication-lift}
Let $C(X,Z)$ be a circuit of size $s$, including its declared inputs, and
let $F(X)=\bigoplus_z C(X,z)$. Given positive unary weights $w_x$, one can
construct a conjunction $H$ of implications on distinct variables with
the following exact certificate family:
\[
 M\cup T\cup\bigcup_{x\in A_0}B_x\quad(A_0\in\Max(F)),
 \qquad V(H)\setminus\{v\}\quad(v\in M\cup T).
\]
Here $|M|=|T|=m=O(s)$, $|B_x|=w_x$, and there is one further variable.
Thus $N=2m+\sum_xw_x+1$ and
\[
 \depth(H)=\min\left\{
 2m+\min_{A_0\in\Max(F)}\sum_{x\in A_0}w_x,\ N-1\right\}.
\]
If all weights are at most $L$, the formula has $O(sL)$ variables and
$O(s^2L)$ clauses. Its implication graph is acyclic of height at most four.
Every coordinate is essential, and the displayed guards are explicit
certificates.
\end{theorem}

\begin{proof}
We begin with the construction for weights equal to one, which has three
tasks. Firstly, we represent $F$ as a parity sum of an implication
formula. Secondly, we add guards that exclude unwanted maximal monomials.
Thirdly, we introduce the weights without leaving the implication
language.

\paragraph{Representing a positive clause by implications.}
We start with the positive compiler $Q_0(X,M_0)$ of \cref{lem:compiler}.
For each clause $a\vee b$, we introduce a private variable $e$ and replace
the clause by
\[
 (a\to e)\wedge(b\to e).
\]
If $a=b=0$, either value of $e$ is allowed, which gives even parity, and
otherwise $e$ must be one, which gives odd parity. Summing over $e$
therefore recovers $a\vee b$. Since different clauses have different
private variables, summing over all of them recovers $Q_0$.

In the resulting formula $Q$, every implication goes from an old
variable to one of the new private variables. We call the old variables
sources and the private variables sinks. Note that all public variables
are sources, and that the variables we must sum over are the old set
$M_0$ together with all sinks. We will need an odd number, at least
three, of each kind of mandatory variable. The number of sources already
has this property. To adjust the number of sinks, we append disjoint
components consisting of two new sources pointing to one new sink, with
all three variables mandatory. Each component has five satisfying
assignments, one when the sink is zero and four when it is one, so its
parity is one and it does not change $F$. Since each component adds two
sources and one sink, at most three such components suffice.

Let $A$ be the set of mandatory sources and $B$ the set of mandatory
sinks, and put $M=A\sqcup B$. We now have $|A|,|B|\ge3$, both odd, and
$\D_MQ=F$.

\paragraph{Adding the guards.}
Let $Y$ contain all variables of $Q$, together with one extra variable
$d$ that $Q$ does not use. For every $i\in M$, we introduce a marker
$t_i$, whose implications constrain every coordinate of $Y$ except $i$:
\[
 \begin{cases}
 t_i\to y &\text{for all }y\in Y\setminus\{i\},\quad i\in A,\\
 y\to t_i &\text{for all }y\in Y\setminus\{i\},\quad i\in B.
 \end{cases}
\]
We conjoin these implications with $Q$ to obtain $H$. For a source
marker, the value $t_i=1$ forces all other old variables to one, which
satisfies $Q$ because every sink is one. For a sink marker, the value
$t_i=0$ forces all other old variables to zero, which also satisfies $Q$,
because every source is zero. Notice that the direction of the
implications is chosen so that both of these assignments are safe.

To treat the two kinds of marker together, we write $u_i=1+t_i$ for
$i\in A$ and $u_i=t_i$ for $i\in B$, since complementing coordinates does
not change certificate sets. In these coordinates, $u_i=0$ means that
the marker is active. With no active marker, $H$ is just $Q$, and one
active marker leaves only its paired old variable unconstrained. Two or
more active markers of the same kind force every old variable to the
same extreme value. Active markers of different kinds are impossible,
because they would force $d$ to be both zero and one, and this is the
role of $d$ in the construction.

Next, we translate these cases into a polynomial identity. We define
\[
 P^+=\prod_{y\in Y}y,\qquad P^-=\prod_{y\in Y}(1+y),\qquad
 P_i^+=\prod_{y\in Y\setminus\{i\}}y,\qquad
 P_i^-=\prod_{y\in Y\setminus\{i\}}(1+y).
\]
These are the indicators of the extreme assignments just described. For
$J\subseteq M$, we put $u_J=\prod_{i\in J}u_i$ and write
$u_{-i}=u_{M\setminus\{i\}}$. The case with no active marker contributes
$u_MQ$. The case with only marker $i$ active contributes
$(1+u_i)u_{-i}P_i^+$ if $i\in A$, and the analogous expression with
$P_i^-$ if $i\in B$. Further, the indicator of at least two active source
markers and no active sink marker is
\[
 u_B\left(1+u_A+\sum_{i\in A}(1+u_i)u_{A\setminus\{i\}}\right)
 =u_B+\sum_{i\in A}u_{-i}.
\]
The first expression removes, from all assignments to these markers, the
cases with zero or one active source marker, and the equality uses that
$|A|$ is odd. The calculation for sink markers is identical. Adding all
cases gives, in the complemented marker coordinates,
\begin{align}
 H={}&u_M\left(Q+\sum_{i\in A}P_i^+
                         +\sum_{i\in B}P_i^-\right)\notag\\
 &+\sum_{i\in A}u_{-i}(P_i^++P^+)
   +\sum_{i\in B}u_{-i}(P_i^-+P^-)\notag\\
 &+u_BP^++u_AP^- .\label{eq:implication-crown}
\end{align}

\paragraph{The maximal monomials.}
The full support has coefficient zero, since $Q$ omits $d$, every
$P_i^\pm$ omits $i$, and every other summand omits a marker. On the
other hand, each support missing just one coordinate of $M\cup T$ has
coefficient one. If the missing coordinate is an old variable $i$, its
term comes from $u_MP_i^\pm$, because any other $P_j^\pm$ omits $j$ and
$Q$ omits $d$. If the missing coordinate is the marker of $i$, its term
comes from $u_{-i}P^\pm$, and the last two summands of
\eqref{eq:implication-crown} omit at least three markers, so they cannot
cancel it. These $2m$ supports of size $N-1$ are therefore maximal, and
they are the guards of the theorem.

Every term outside $u_MQ$ misses an old mandatory variable or a marker,
and is therefore contained in a guard. The same is true of every term of
$u_MQ$ that misses an old mandatory variable. We are left with the terms
of $u_MQ$ that contain all of $M\cup T$. No guard contains them, and no
other summand can cancel them. Moreover, none of them contains a guard,
since they omit $d$. Their maximality is consequently decided entirely
within $Q$.

The identity $\D_MQ=F$ says exactly that the supports of $Q$ containing
$M$ are the sets $M\cup A_0$ for the supports $A_0$ of $F$. Hence the
remaining maximal supports of $H$ are $M\cup T\cup A_0$, with
$A_0\in\Max(F)$, and this also covers $F=0$ and $F=1$. Since the guards
together contain every coordinate, every coordinate occurs in the ANF
and is essential.

\paragraph{Assigning weights.}
Before adding the markers, we replace each public source $x$ by the OR
of $w_x$ new variables, which form a block $B_x$. This substitution stays
within implication formulas, because an arc $x\to e$ is replaced by the
arcs $b\to e$ with $b\in B_x$, so the number of clauses grows only
linearly with the size of the block.

To see that this substitution gives the intended cost, recall that the
ANF of an OR contains every nonempty monomial on its block. Each
monomial obtained by substitution identifies the original support
through the blocks it meets, so different original supports cannot
cancel. Within one block, any partial support can be enlarged to the
whole block and still occurs. It follows that maximal supports use whole
blocks, and correspond exactly to the old maximal supports. The
substitution commutes with $\D_M$, since the public blocks are disjoint
from $M$, and the mandatory sources and sinks do not change. The guard
argument therefore applies to the substituted compiler, and proves the
claimed family.

There are $O(sL)$ old variables and clauses, and each of the $m=O(s)$
markers adds one implication for each old variable except its partner,
which gives $O(s^2L)$ clauses in total. All arcs go forward through four
layers: source markers, old sources together with $d$, old sinks, and
sink markers. Some arcs skip a layer, but none of them goes backward, so
the implication graph is acyclic of height at most four. The counts of
variables and the formula for $\depth(H)$ now follow from the displayed
family of certificates. As with every implication formula, both the all
zero and the all one assignments satisfy $H$.
\end{proof}

\subsection{Proof of the dichotomy}
\begin{proof}[Proof of \cref{thm:binary}]
Suppose first that every allowed relation is affine. Combining their
linear equations gives a system $Hx=b$ over $\F$. If it is inconsistent,
the represented function is zero and has no certificate, so we assume
from now on that it is consistent.

We choose a set $S$ of variables, and fix the other variables to $y$.
The remaining system is
\[
 H_Sx_S=b+H_{V\setminus S}y.
\]
This system has either no solution or exactly
$2^{|S|-\operatorname{rank}H_S}$ of them. For every fiber to have odd
parity, two conditions must therefore hold. Firstly, the columns in $S$
must be independent. Secondly, every right hand side in the display must
lie in their span. Taking $y=0$, and then each unit vector, shows that
the second condition requires $b$ and all columns outside $S$ to lie in
that span, so $S$ must be a column basis of $H$. Conversely, a column
basis works: consistency places $b$ in the column span of $H$, and each
remaining system then has exactly one solution. The certificates are
therefore exactly the nonempty column bases, and Gaussian elimination
gives $\depth=\operatorname{rank}H$ when the rank is positive. At rank
zero the function is constant one, and the requirement of a nonempty set
leaves no certificate.

Next, we identify all remaining languages. Every unary relation is
affine, including the empty relation, which is described by an
inconsistent system. A binary relation with zero, one, two or four
tuples is also affine, since two distinct vectors over $\F$ form an
affine line. A binary relation that is not affine must therefore contain
exactly three tuples. Depending on which tuple it excludes, it is OR,
NAND, implication, or reversed implication.

Hardness for OR is \cref{thm:hardness}, and complementing every variable
gives hardness for NAND. For implication, we apply
\cref{thm:implication-lift} to the selector
construction~\eqref{eq:selector}, giving weight one to its paired
assignment variables and weight $L$ to its universal variables. This
yields the same gap as before, $K$ against at least $K+L$. Reversing the
two arguments handles reversed implication. Note that these reductions
use no constants, since every new variable is an input of the
constructed formula, and that membership in $\Sig$ is proved as for BCD.

Finally, any unary or binary relation can be expressed by a signed
$2$CNF of constant size, with one clause for each forbidden tuple. This
changes the input size by only a constant factor, so
\cref{thm:small-one} supplies the stated parameterized algorithm.
\end{proof}

The implication construction also preserves the two approximation bounds
of \cref{thm:hardness}, since it has the same $O(s^2L)$ bound on clauses.
It has a useful interpretation in terms of orders as well. The
coordinates set to one form an upset of the partial order defined by the
acyclic implication graph, so the hardness holds for upsets of posets of
height four. This depends, however, on how the poset is encoded. A DAG
encoding, or its transitive reduction, retains the size bound used in
the approximation proof, whereas writing out the full matrix of the
transitive closure may not.

\section{The number of quadratic rounds}\label{app:rounds}
In this appendix, we prove \cref{thm:rounds}. The upper bound follows
from the degree of the computed function. The lower bound needs more
care, since a SAT instance must be represented using few quadratic
rounds, while keeping the number of inputs small enough that an exponent
of $2^{o(r)}$ would place NP in SUBEXP.

\begin{proof}[Proof of \cref{thm:rounds}]
We begin with the algorithm. After $j$ rounds, every coordinate has
Boolean degree at most $2^j$, because a quadratic map multiplies at most
two coordinates at a time, and reducing $x_i^2$ to $x_i$ can only
decrease the degree. We can therefore compute each coordinate by storing
its table of multilinear coefficients. To multiply two tables, we take
the union of each pair of supports and add the resulting coefficients
modulo two.

At the last round, we list the nonzero supports and discard every
support that is contained in a larger one. By \cref{lem:anf}, the
smallest remaining nonempty support is a smallest certificate, and if
none remains, there is no certificate. Since the tables contain at most
$n^{O(2^r)}$ entries, even pairwise multiplication and containment tests
fit within $n^{O(2^r)}\poly(\ell)$ time.

\paragraph{Encoding assignments by monomials.}
For the lower bound, we start with a $3$SAT instance on $N$ variables. We
remove duplicate and tautological clauses, and decide empty clauses, as
well as instances with fewer than two remaining clauses, directly. We
then pad the clause list by duplication until its length is a power of
two, say $m=2^q$. Since there are only $O(N^3)$ distinct clauses of width
at most three, we have $m=O(N^3)$ and $q=O(\log N)$.

Fix the constant $c>0$ of the theorem. We divide the variables into
$p=\lceil N/b\rceil$ groups, where $b=\lceil c\log_2N\rceil$, and pad the
last group with unused SAT variables if needed. For every group $i$ and
every assignment $a\in\{0,1\}^b$ to that group, we introduce an input
variable $x_{i,a}$. Selecting one such variable from each group then
describes a full SAT assignment. Note that the added unused SAT variables
are ordinary assignment choices, and that we never sum over them as
parity variables.

For each group, we define
\[
 T_i=\sum_a x_{i,a},\qquad
 L_{i,j}=\sum_{\substack{a\text{ satisfies no literal of}\\
                         \text{clause }j\text{ in group }i}}x_{i,a}.
\]
The product $\prod_iT_i$ has one monomial for every full assignment, and
the product $\prod_iL_{i,j}$ keeps exactly those monomials whose
assignments falsify clause $j$. A group not used by that clause imposes
no restriction, so its factor is simply $L_{i,j}=T_i$.

We would like to test each clause separately, because adding these
products over all clauses would let two false clauses cancel each other.
To do so, we introduce $q$ address inputs $y$, and let $\delta_j(y)$ be
the indicator of address $j$, i.e., the product of the $q$ literals that
specify that address. Consider
\begin{equation}\label{eq:round-polynomial}
 F(x,y)=\prod_{i=1}^pT_i+
        \sum_{j=1}^m\delta_j(y)\prod_{i=1}^pL_{i,j}.
\end{equation}
At a fixed address $j$, the second summand removes the monomials of the
assignments that falsify clause $j$. Writing $A_\alpha$ for the $p$
inputs that encode a full assignment $\alpha$, we therefore have
\[
 F(x,y)=\sum_\alpha x_{A_\alpha}h_\alpha(y),
\]
where $h_\alpha(y)$ is one exactly when $\alpha$ satisfies the addressed
clause. In particular, $h_\alpha\equiv1$ exactly when $\alpha$ satisfies
every clause.

The sets $A_\alpha$ are pairwise incomparable, because each of them
chooses exactly one input from each group. A monomial associated with
one assignment can therefore never contain a monomial associated with
another assignment, even after address variables are added. Within the
terms for a fixed $\alpha$, the maximal supports are $A_\alpha$ together
with a maximal support of $h_\alpha$, and they have size at most $p$
exactly when $h_\alpha\equiv1$. If $h_\alpha=0$, that assignment
contributes no term at all. We conclude that
\[
 \depth(F)\le p
 \quad\Longleftrightarrow\quad
 \text{the original SAT instance is satisfiable}.
\]

\paragraph{Computing the polynomial in few rounds.}
One affine linear map forms all $T_i$, all $L_{i,j}$, the address
literals, and the constant one. Next, we compute all products in
\eqref{eq:round-polynomial} by binary product trees, in which the group
products have $p$ factors and the address indicators have $q$ factors.
Padding with factors equal to one lets all of these trees run in
parallel for
\[
 d=\max\{\lceil\log_2p\rceil,\lceil\log_2q\rceil\}
\]
rounds. One final quadratic coordinate forms the sum in
\eqref{eq:round-polynomial}, where each term multiplies a computed
address indicator by a computed group product. This last step is
permitted, because a coordinate of a quadratic map may sum many
quadratic terms. The total number of rounds is $r=d+2$.

For sufficiently large $N$ we have $q\le p$, and the number of initial
inputs and the number of rounds then satisfy
\[
 n=p2^b+q=\Theta(N^{c+1}/\log N),\qquad
 2^r=\Theta(N/\log N).
\]
All maps together have description length $\ell=N^{O_c(1)}$. Now
suppose that an algorithm runs in time $n^{a(r)}\ell^{O(1)}$ with
$a(r)=2^{o(r)}$. Since $2^r\le N$ for large $N$, we have
$a(r)=2^{o(\log N)}=N^{o(1)}$, and further $\log n=O_c(\log N)$. On this
construction, the running time would therefore be
\[
 n^{a(r)}\ell^{O(1)}=2^{a(r)\log_2n}\,N^{O_c(1)}=2^{N^{o(1)}}.
\]
Then $3$SAT would be in SUBEXP, and so would every problem in NP.
Eliminating $N$ from the displayed counts of inputs and rounds yields
\[
 2^r=\Theta\!\left(
 n^{1/(c+1)}/(\log n)^{c/(c+1)}\right),
\]
which is the claimed range.
\end{proof}

We note that the restriction to this range matters. Once $2^r$ is
comparable to $n$ or larger, computing the full truth table can already
improve on the bound based on degree. The reduction also does not rule
out a running time with an arbitrary additional factor that depends only
on $r$.

\section{Ordered branching programs}\label{app:branching}
In this appendix, we prove \cref{thm:branching} and its lower bound. A
fixed variable order gives another way to keep the search small: we read
one variable at a time, and decide whether to include it in the cube.
For the derivative to be the constant one, the coefficient of the cube's
own monomial must be one, and the coefficients of all larger supports
must be zero. Note that we do not need to store those larger supports
separately, because their coefficient computations can be represented by
a vector space.

\begin{theorem}\label{thm:branching}
Given
\[
 f(x)=u^{\mathsf T}\prod_{i=1}^{n}(A_i+x_iB_i)v
\]
over $\F$, with matrices of dimension at most $w$ and each variable appearing
in one layer, one can compute $\depth(f)$ and a minimizing certificate, or
report infeasibility, in time $n2^{O(w^2)}$. The same bound holds for
$\zeta(f)$.
\end{theorem}

\subsection{Keeping track of larger monomials}
\begin{proof}
We pad the matrix dimensions to $w$. To obtain the coefficient of $x_S$,
we choose $B_i$ at each layer $i\in S$ and $A_i$ at every other layer,
and then multiply on the left by $u^{\mathsf T}$ and on the right by $v$.

After the first $j$ layers, the chosen matrices give a row vector, which
the remaining layers will turn into a coefficient. For a fixed selected
set $S\subseteq[j]$, we keep two pieces of information, namely the row
for $S$ itself and the span of the rows for all of its strict supersets
in this prefix. Formally, we put $M_i(U)=B_i$ if $i\in U$ and
$M_i(U)=A_i$ otherwise, and define
\[
 p=u^{\mathsf T}\prod_{i\le j}M_i(S),\qquad
 W=\spanf\left\{u^{\mathsf T}\prod_{i\le j}M_i(S\cup R):
       \varnothing\ne R\subseteq[j]\setminus S\right\}.
\]
If there is no strict superset, $W$ is the zero space. The span
suffices, because testing whether all of these rows eventually give zero
is a linear condition.

A state of the dynamic program is a triple $(p,W,b)$, where the bit $b$
says whether $S$ is nonempty, and we store the least size of a selection
that produces each state. Initially, the state is
$(u^{\mathsf T},\{0\},0)$ at cost zero. At the next layer, we write its
two matrices as $A,B$.

If we select the new variable, every relevant monomial must use it.
Hence we multiply both the exact row and the rows of strict supersets by
$B$:
\[
 (p,W)\longmapsto(pB,WB).
\]
The cost increases by one, and the bit becomes true.

If we skip the new variable, the exact row becomes $pA$. A strict
superset of the new selection then has three possibilities: it adds old
variables only, it adds the new variable as well as old variables, or it
adds only the new variable. The corresponding rows belong to $WA$, $WB$
and $\{pB\}$ respectively. Thus the transition for a skipped variable is
\[
 (p,W)\longmapsto
 \bigl(pA,\spanf(WA\cup WB\cup\{pB\})\bigr),
\]
with unchanged cost and bit. Note that both transitions preserve the
definitions of $p$ and $W$.

After the last layer, multiplying by $v$ gives the actual coefficients.
In particular, $pv$ is the coefficient of $x_S$, and every coefficient
of a strict superset vanishes exactly when every row in $W$ is
orthogonal to $v$. By \cref{lem:anf}, we accept a certificate exactly
when
\[
 b=1,\qquad pv=1,\qquad W\subseteq v^\perp.
\]
For a zero cube, we change $pv=1$ to $pv=0$, which requires all
coefficients whose supports contain $S$, including $S$ itself, to
vanish. Since equal states have the same transitions and acceptance
conditions under every continuation, retaining only the cheapest
selection is valid. Stored predecessor choices recover a minimizing set.

There are $2^w$ possible rows and at most $2^{w^2}$ subspaces of $\F^w$,
since a basis of at most $w$ rows specifies any such space. We use bases
in reduced row echelon form to compare spaces, and to compute the
transitions in time polynomial in $w$. Including the bit and all $n$
layers gives $n2^{O(w^2)}$ time.

Restrictions on the selection are accommodated by the same algorithm. We
disable the selected transition at forbidden layers, and force it, with
the prescribed cost, at mandatory layers. When mandatory variables
represent an initial parity projection, they do not set the bit, which
records nonemptiness among the eligible inputs only.
\end{proof}

This includes complete layered OBDDs. If $T_i^0,T_i^1$ are the
transition matrices for the two values of variable $i$, we take
\[
 A_i=T_i^0,\qquad B_i=T_i^0+T_i^1.
\]
Then $A_i+x_iB_i$ is $T_i^0$ at $x_i=0$ and $T_i^1$ at $x_i=1$, and the
width is the number of states in a complete layer. An OBDD that skips
tests must first be completed, and our bound uses its width after
completion. Note also that the verifier in \cref{lem:one-check} is
another instance of the same ordered matrix calculation.

\subsection{A lower bound for OBDD width}
We now show that a running time of $2^{w^{o(1)}}\poly(\ell)$ is
impossible, unless $\mathrm{NP}\subseteq\mathrm{SUBEXP}$.

\begin{proposition}\label{prop:obdd-hard}
BCD is NP-complete on complete deterministic OBDDs. Unless
$\mathrm{NP}\subseteq\mathrm{SUBEXP}$, it has no
$2^{w^{o(1)}}\poly(\ell)$ algorithm in their layered width $w$.
\end{proposition}
\begin{proof}
For a prescribed set, we follow only its chosen transitions in the
preceding algorithm. There is one row and one subspace to update at each
layer, so verification takes polynomial time, and guessing the set
proves membership in NP.

For hardness, we start with a $3$SAT formula on $n$ variables and
$m\ge2$ nonempty clauses, and handle the trivial cases directly. We
duplicate clauses to obtain $m'=2^q<2m$ clauses, indexed by address bits
$y$, and use the same Boolean function as in \eqref{eq:ethcircuit}:
\[
 F(a,b,y)=\prod_i(a_i+b_i)
       +\prod_i\bigl(a_i g_i^1(y)+b_i g_i^0(y)\bigr),
\]
where $g_i^b(y)$ is zero if the value $b$ for source variable $i$
satisfies a literal of the addressed clause, and one otherwise. As shown
in the proof of \cref{thm:eth},
\[
 F=\sum_\alpha x_{A_\alpha}h_\alpha(y).
\]
Here $A_\alpha$ chooses one variable from each pair $a_i,b_i$, and
$h_\alpha$ indicates that the addressed clause is satisfied. Since the
different $A_\alpha$ are incomparable and have size $n$, a maximal
support of size at most $n$ exists exactly when some $h_\alpha$ is
identically one. Thus $\depth(F)\le n$ exactly when the source formula is
satisfiable.

To compute $F$ with an OBDD, we read the address bits first and retain
the clause index. We then read $a_1,b_1,\ldots,a_n,b_n$, keeping, besides
the index, two accumulator bits $P,R$ that are initially one. After
reading pair $i$, we update
\[
 P\leftarrow P(a_i+b_i),\qquad
 R\leftarrow R(a_i g_i^1(y)+b_i g_i^0(y)).
\]
While reading the second member of a pair, we also retain the value of
the first member, and at the end we output $P+R$. These transitions are
deterministic and use at most $8m'$ states in each layer. Since the
number of layers and the total size of the OBDD are polynomial, this
proves NP-hardness.

For the lower bound, note that the width is $w\le8m'<16m$, which is at
most linear in the length $\lambda$ of the source formula. A
$2^{w^{o(1)}}\poly(\ell)$ algorithm would therefore decide $3$SAT in time
$2^{\lambda^{o(1)}}$, and $3$SAT, together with every problem in NP,
would then be in SUBEXP.
\end{proof}

We acknowledge, however, that the bounds leave a wide gap: the algorithm
has an exponent quadratic in $w$, whereas the lower bound only excludes
an exponent of $w^{o(1)}$. Note also that the fixed read-once order is
important for the algorithm. For example, $(1+x)x=0$ as a Boolean
polynomial, whereas treating its two occurrences of $x$ as independent
variables would give a nonzero polynomial and a false certificate. The
argument above therefore does not extend to programs that read a
variable more than once.

\section{Hardness for strongly collapsible independence complexes}\label{app:topology}
In this appendix, we prove the last remark of the introduction. A
positive $2$CNF can be viewed as a graph in which each clause
$x_u\vee x_v$ becomes an edge $uv$, and its satisfying assignments are
then the vertex covers of the graph. It is natural to ask whether the
hardness comes from a complicated independence complex. We show that it
persists even when this complex can be reduced to a point by a
particularly simple kind of collapse.

For a graph $G$, we write
\[
 f_G=\bigwedge_{uv\in E(G)}(x_u\vee x_v).
\]
The independence complex $\Ind(G)$ consists of all independent vertex
sets, including the empty set. A vertex of a simplicial complex is
dominated by another if every maximal face containing the first also
contains the second, and deleting dominated vertices is called a strong
collapse~\cite{barmak2012}.

The following extension has two useful features: its effect on the
smallest certificate is known exactly, and its independence complex has
an explicit collapse sequence.

\begin{lemma}\label{lem:wrapper}
Suppose $f_G$ is nonconstant and its full ANF coefficient is zero. Add a
vertex $u$ adjacent to all of $G$, forming $J$. Add $a,c,d$, join both $a$
and $c$ to every vertex of $J$, and add the edge $cd$. The resulting graph
$H$ is connected, every coordinate of $f_H$ is essential, and
\[
 \depth(f_H)=\depth(f_G)+3.
\]
Moreover, $\Ind(H)$ has an explicit strong collapse to a point.
\end{lemma}
\begin{proof}
We first compute the change in the size of a certificate. Let
$X=\prod_{v\in V(G)}x_v$. In the first extension, setting $u=1$ leaves
$f_G$, whereas setting $u=0$ forces every old variable to one. Thus
\[
 f_J=uf_G+X+uX.
\]
In the second extension, if some coordinate of $J$ is zero, its edges to
$a$ and $c$ force both of these variables to one, and the edge $cd$ is
then satisfied regardless of $d$. If all coordinates of $J$ are one, only
the constraint $c\vee d$ remains. Since the product of the coordinates of
$J$ is $uX$, these two cases give
\[
 f_H=ac f_J+uX(c+d+ac+cd)
     =acu f_G+acX+cuX+duX+cduX.
\]

The last expression makes the maximal supports explicit. The terms $cuX$
and $duX$ are contained in $cduX$, so neither of them is maximal. The
terms from $acu f_G$ contain both $a$ and $u$, and neither remaining
displayed term contains both. Also, the full old support $X$ is absent
from $f_G$. It follows that the maximal supports are exactly
\[
 \{a,c,u\}\cup S\quad(S\in\Max(f_G)),\qquad
 V(G)\cup\{a,c\},\qquad V(G)\cup\{c,d,u\}.
\]
Writing $n=|V(G)|$, their smallest size is
\[
 \min\{\depth(f_G)+3,n+2,n+3\}=\depth(f_G)+3.
\]
For the equality, note that a nonconstant polynomial has a nonempty
maximal support, and the absent full term makes its size at most $n-1$.
The last two displayed supports together contain every coordinate, so
every coordinate occurs in the ANF and is essential. The added edges also
make $H$ connected.

Now consider the collapse. The neighbors of $a$ are all the vertices of
$J$, whereas the neighbors of $c$ are those same vertices together with
$d$, and $a,c$ are not adjacent. Any independent set containing $c$ can
therefore include $a$ as well. In particular, every maximal such set
already contains $a$, so $c$ is dominated by $a$ in $\Ind(H)$, and we
delete $c$. The vertex $d$ is now isolated in the remaining graph. Since
every maximal independent set contains it, it dominates every other
vertex of the remaining independence complex. Deleting those vertices one
by one leaves only $d$, which is the promised strong collapse.
\end{proof}

Every output of the crown in \cref{lem:crown} satisfies the hypotheses of
\cref{lem:wrapper}, since it is nonconstant and its full ANF coefficient
is zero. The extension adds four vertices, $O(N)$ edges, and exactly
three to the optimum. Applying it to the gap construction of
\cref{thm:hardness} therefore preserves completeness and the nearly
linear approximation lower bounds, and the collapse sequence can be
supplied with the instance. Thus a simple topology for the whole
independence complex does not make the certificate problem easy.

\section*{AI methodology}
AI assistants (ChatGPT and Claude) were used to explore constructions,
develop and review proof arguments, check related work and write finite
verification programs. The computational checks compare finite truth tables, ANF supports, and algorithm outputs; they are supplementary to the proofs.

\newcommand{\etalchar}[1]{$^{#1}$}

\end{document}